# Functional Splicing Network Reveals Extensive Regulatory Potential of the Core Spliceosomal Machinery

Panagiotis Papasaikas,[1,2,4] J. Ramón Tejedor,[1,2,4] Luisa Vigevani,[1,2] and Juan Valcárcel[1,2,3,*]

[1]Centre de Regulació Genòmica, Dr. Aiguader 88, 08003 Barcelona, Spain
[2]Universitat Pompeu Fabra, Dr. Aiguader 88, 08003 Barcelona, Spain
[3]Institució Catalana de Recerca i Estudis Avançats, Passeig Lluis Companys 23, 08010 Barcelona, Spain
[4]Co-first authors

## SUMMARY

**Pre-mRNA splicing relies on the poorly understood dynamic interplay between >150 protein compo- nents of the spliceosome. The steps at which splicing can be regulated remain largely unknown. We sys- tematically analyzed the effect of knocking down the components of the splicing machinery on alterna- tive splicing events relevant for cell proliferation and apoptosis and used this information to reconstruct a network of functional interactions. The network accurately captures known physical and functional associations and identifies new ones, revealing remarkable regulatory potential of core spliceosomal components, related to the order and duration of their recruitment during spliceosome assembly. In contrast with standard models of regulation at early steps of splice site recognition, factors involved in catalytic activation of the spliceosome display regu- latory properties. The network also sheds light on the antagonism between hnRNP C and U2AF, and on tar- gets of antitumor drugs, and can be widely used to identify mechanisms of splicing regulation.**

## INTRODUCTION

Pre-mRNA splicing is carried out by the spliceosome, one of the most complex molecular machineries of the cell, composed of five small nuclear ribonucleoprotein particles (U1, U2, U4/5/6 snRNP) and about 150 additional polypeptides (reviewed by Wahl et al., 2009). Detailed biochemical studies using a small number of model introns have delineated a sequential pathway for the assembly of spliceosomal subcomplexes. For example, U1 snRNP recognizes the 5' splice site, and U2AF—a hetero- dimer of 35 and 65 kDa subunits—recognizes sequences at the 3' end of introns. U2AF binding helps to recruit U2 snRNP to the upstream branchpoint sequence, forming complex A. U2 snRNP binding involves interactions of pre-mRNA sequences with U2 snRNA as well as with U2 proteins (e.g., SF3B1). Subse- quent binding of preassembled U4/5/6 tri-snRNP forms complex B, which after a series of conformational changes forms com- plexes Bact and C, concomitant with the activation of the two catalytic steps that generate splicing intermediates and prod- ucts. Transition between spliceosomal subcomplexes involves profound dynamic changes in protein composition as well as extensive rearrangements of base-pairing interactions between snRNAs and between snRNAs and splice site sequences (Wahl et al., 2009). RNA structures contributed by base-pairing interac- tions between U2 and U6 snRNAs serve to coordinate metal ions critical for splicing catalysis (Fica et al., 2013), implying that the spliceosome is an RNA enzyme whose catalytic center is only established upon assembly of its individual components.

Differential selection of alternative splice sites in a pre-mRNA (alternative splicing, AS) is a prevalent mode of gene regulation in multicellular organisms, often subject to developmental regula- tion (Nilsen and Graveley, 2010) and frequently altered in disease (Cooper et al., 2009; Bonnal et al., 2012). Substantial efforts made to dissect mechanisms of AS regulation on a relatively small number of pre-mRNAs have provided a consensus picture in which protein factors recognizing cognate auxiliary sequences in the pre-mRNA promote or inhibit early events in spliceosome assembly (Fu and Ares, 2014). These regulatory factors include members of the hnRNP and SR protein families, which often display cooperative or antagonistic functions depending on the position of their binding sites relative to the regulated splice sites. Despite important progress (Barash et al., 2010; Zhang et al., 2010), the combinatorial nature of these contributions compli- cates the formulation of integrative models for AS regulation.

To systematically capture the contribution of splicing regulato- ry factors, including components of the core spliceosome (Clark et al., 2002; Park et al., 2004; Pleiss et al., 2007; Saltzman et al., 2011) to AS regulation, we set up a high-throughput screen to evaluate the effects of knocking down each individual splicing component or regulator on 36 functionally important AS events (ASEs) and developed a framework for data analysis and network modeling. Network-based approaches can provide rich representations for real-world systems that capture their or- ganization more adequately than more traditional approaches or mere cataloguing of pairwise relationships. Numerous studies have utilized them as a platform for deriving comprehensive tran- scriptional, metabolic, and physical interaction maps in several

**A**

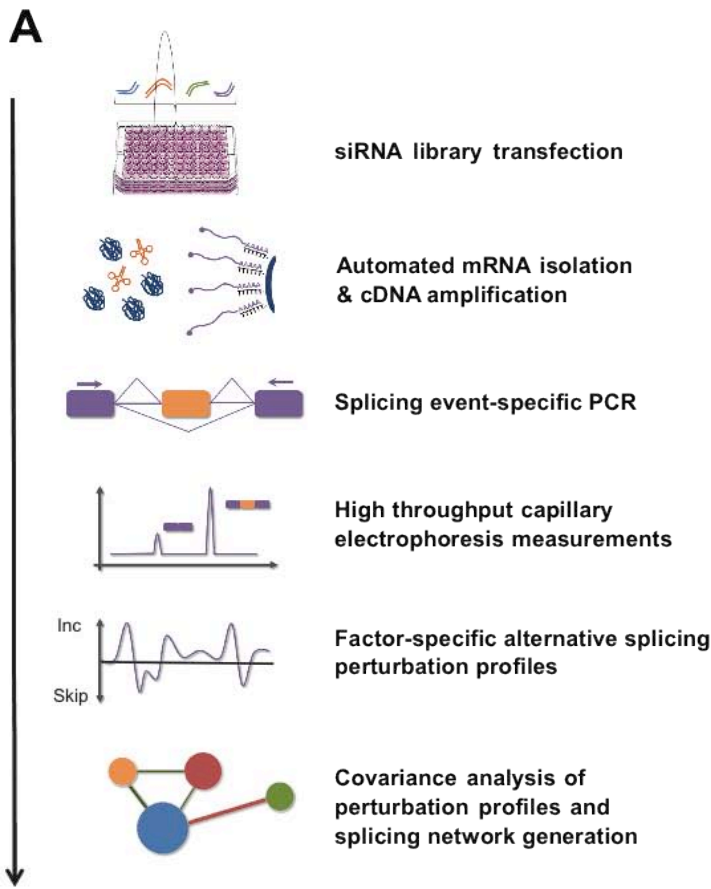

siRNA library transfection

Automated mRNA isolation & cDNA amplification

Splicing event-specific PCR

High throughput capillary electrophoresis measurements

Factor-specific alternative splicing perturbation profiles

Inc

Skip

Covariance analysis of perturbation profiles and splicing network generation

**B**

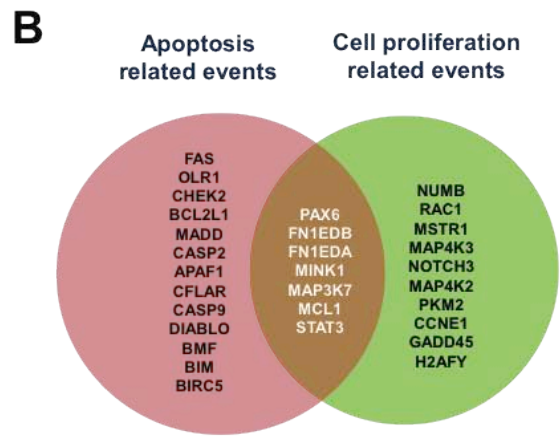

Apoptosis related events

Cell proliferation related events

FAS
OLR1
CHEK2
BCL2L1
MADD
CASP2
APAF1
CFLAR
CASP9
DIABLO
BMF
BIM
BIRC5

PAX6
FN1EDB
FN1EDA
MINK1
MAP3K7
MCL1
STAT3

NUMB
RAC1
MSTR1
MAP4K3
NOTCH3
MAP4K2
PKM2
CCNE1
GADD45
H2AFY

**C**

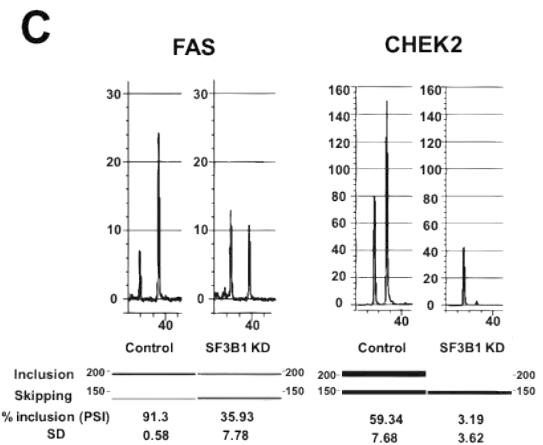

FAS        CHEK2

Control   SF3B1 KD       Control   SF3B1 KD

| | Inclusion 200 Skipping 150 | | |
|---|---|---|---|
| % inclusion (PSI) | 91.3 | 35.93 | 59.34 | 3.19 |
| SD | 0.58 | 7.78 | 7.68 | 3.62 |

**D**

ΔPSI

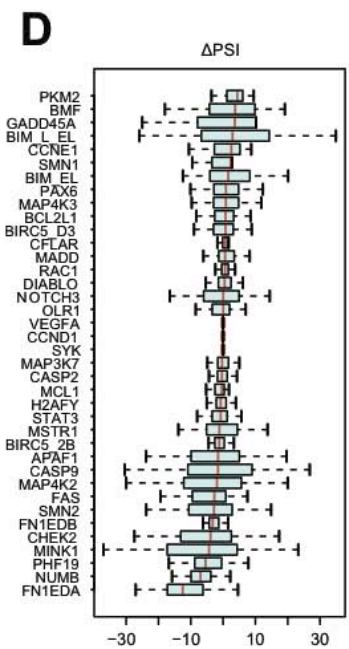

PKM2
BMF
GADD45A
BIM_L_EL
CCNE1
SMN1
BIM_EL
PAX6
MAP4K3
BCL2L1
BIRC5_D3
CFLAR
MADD
RAC1
DIABLO
NOTCH3
OLR1
VEGFA
CCND1
SYK
MAP3K7
CASP2
MCL1
H2AFY
STAT3
MSTR1
BIRC5_2B
APAF1
CASP9
MAP4K2
FAS
SMN2
FN1EDB
CHEK2
MINK1
PHF19
NUMB
FN1EDA

-30   -10   10   30

**E**

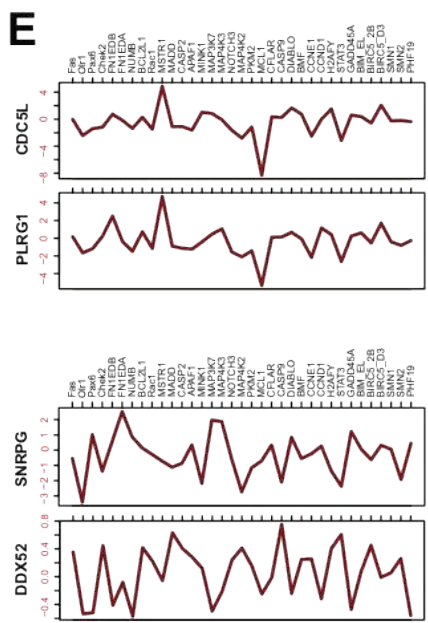

CDC5L

PLRG1

SNRPG

DDX52

**F**

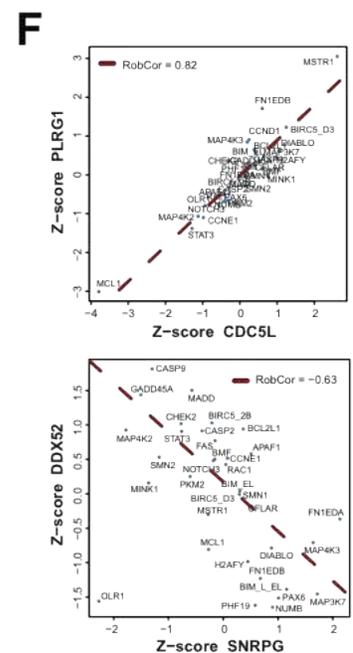

RobCor = 0.82

Z-score PLRG1

Z-score CDC5L

RobCor = -0.63

Z-score DDX52

Z-score SNRPG

(legend on next page)

organisms and contexts (e.g., Chatr-Aryamontri et al., 2013; Franceschini et al., 2013; Wong et al., 2012). Their application has also proven invaluable for deciphering regulatory relationships in transcriptional circuits, for studying their properties upon perturbation, and for connecting physiological responses and diseases to their molecular underpinnings (e.g., Kim et al., 2012; Watson et al., 2013; Yang et al., 2014).

Key to our approach is the premise that the distinct profiles of splicing changes caused by perturbation of regulatory factors depend on the functional bearings of those factors and can, therefore, be used as proxies for inferring functional relationships. We derive a network that recapitulates the topology of known splicing complexes and provides an extensive map of >500 functional associations among ~200 splicing factors (SFs). We use this network to identify general and particular mechanisms of AS regulation and to identify key SFs that mediate the effects on AS of perturbations induced by iron (see accompanying manuscript by Tejedor et al., 2014, in this issue of Molecular Cell) or by antitumor drugs targeting components of the splicing machinery. Our study offers a unique compendium of functional interactions among SFs, a discovery tool for coupling cell-perturbing stimuli to splicing regulation, and an expansive view of the splicing regulatory landscape and its organization.

## RESULTS

Results from a genome-wide siRNA screen to identify regulators of Fas/CD95 AS revealed that knockdown of a significant fraction of core SFs (defined as proteins identified in highly purified spliceosomal complexes assembled on model, single-intron pre-mRNAs; Wahl et al., 2009) caused changes in Fas/CD95 exon 6 inclusion (Tejedor et al., 2014). The number of core factors involved and the extent of their regulatory effects exceeded those of classical splicing regulatory factors like SR proteins or hnRNPs. To systematically evaluate the contribution of different classes of splicing regulators to AS regulation, we set up a screen in which we assessed the effects of knockdown of each individual factor on 38 ASEs relevant for cell proliferation and/or apoptosis (Figures 1A and 1B). The ASEs were selected due to their clear impact on protein function and their documented biological relevance (see Table S1 available online). A library of 270 siRNA pools (Table S2) was used, corresponding to genes encoding core spliceosomal components as well as auxiliary regulatory SFs and factors involved in other RNA-processing steps, including RNA stability, export, or polyadenylation. In addition, 40 genes involved in chromatin structure mod-ulation were included, to probe for possible functional links between chromatin and splicing regulation (Luco et al., 2011).

The siRNA pools were individually transfected in biological triplicates in HeLa cells using a robotized procedure (Supplemental Experimental Procedures). Seventy-two hours posttransfection, RNA was isolated, cDNA libraries generated using oligo-dT and random primers, and the patterns of splicing probed by PCR using primers flanking each of the alternatively spliced regions (Figure 1A). PCR products were resolved by high-throughput capillary electrophoresis (HTCE) and the ratio between isoforms calculated as the percent spliced in (PSI) index (Katz et al., 2010), which represents the percentage of exon inclusion/alternative splice site usage. Knockdowns of factors affecting one ASE (Fas/CD95) were also validated with siRNA pools from a second library and by a second independent technique (Illumina HiSeq sequencing; Tejedor et al., 2014). Conversely, the effects of knocking down core SFs (P14, SNRPG, and SF3B1) on the ASEs were robust when different siRNAs and RNA purification methods were used (Figure S1A). Several pieces of evidence indicate that the changes in AS are not due to the induction of cell death upon depletion of essential SFs. First, cell viability was not generally compromised after 72 hr of knockdown of 13 individual core SFs (Figure S1C). Second, cells detached from the plate were washed away before RNA isolation. Third, splicing changes upon induction of apoptosis with staurosporine were distinct from those observed upon knockdown of SFs (Figure S1D).

Overall the screen generated a total of 33,288 PCR data points. These measurements provide highly robust and sensitive estimates of the relative use of competing splice sites. Figure 1D shows the spread of the effects of knockdown for all the factors in the screen for the ASEs analyzed. Three events (VEGFA, SYK, and CCND1) were excluded from further analysis, because they were not affected in the majority of the knockdowns (median absolute DPSI <1, Figure 1D). From these data we generated, for every knockdown condition, a perturbation profile that reflects its impact across the 35 ASEs in terms of the magnitude and direction of the change. Figures 1E and 1F show examples of such profiles and the relationships between them: in one case, knockdown of CDC5L or PLRG1 generates very similar perturbation profiles (upper diagrams in Figures 1E and 1F), consistent with their well-known physical interactions within the PRP19 complex (Makarova et al., 2004). In contrast, the profiles associated with the knockdown of SNRPG and DDX52 are to a large extent opposite to each other (lower diagrams in Figures 1E and 1F), suggesting antagonistic functions. Perturbation profiles and relationships between factors were not significantly changed

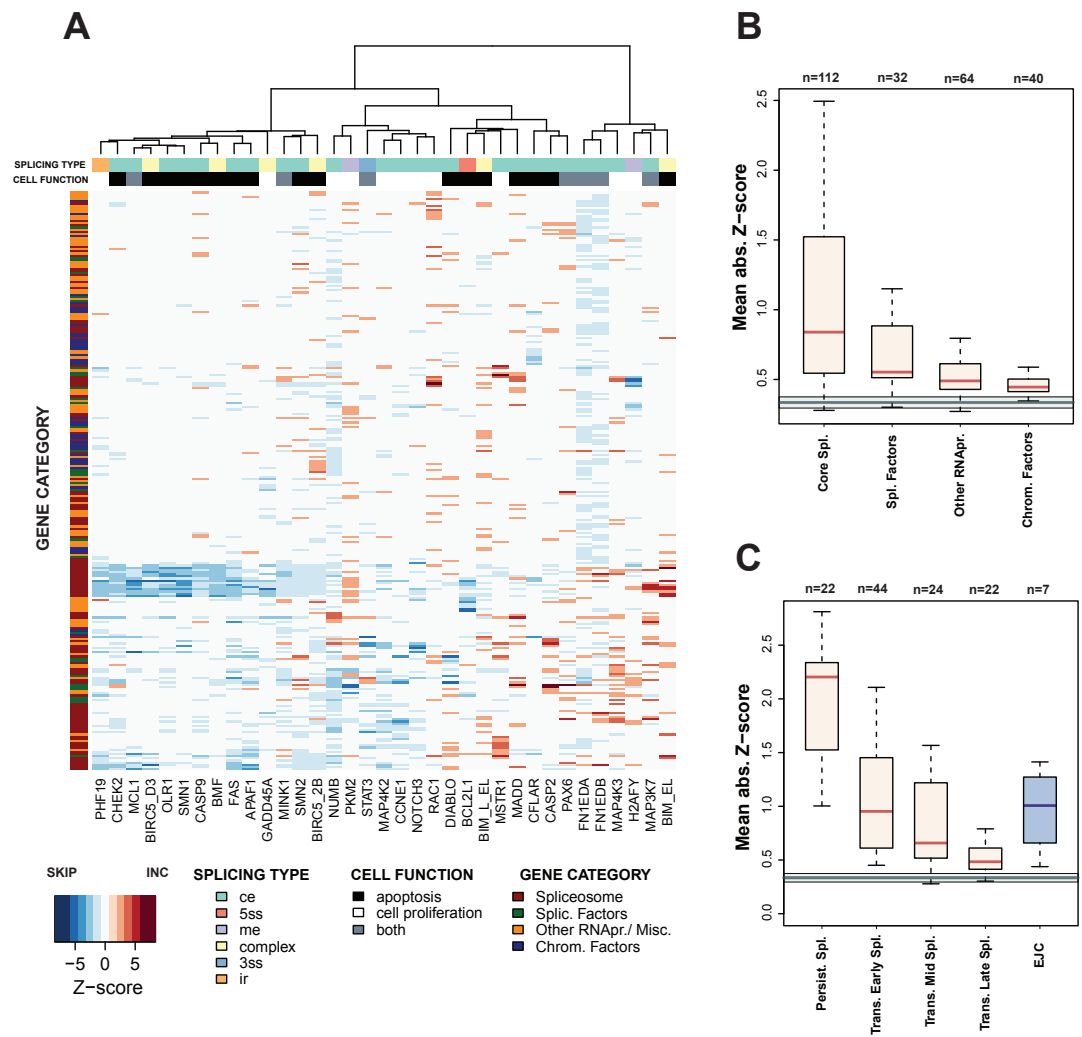

**Figure 2. Coordinated Regulation of Splicing Events by Coherent Subsets of SFs**

(A) Heatmap representation of the results of the screen. ASEs used in this study are on the x axis, while knockdown conditions on the y axis. Data are clustered on both dimensions (ward linkage, similarity measure based on Pearson correlation). Information about ASEs, (a) type (ce, cassette exon; 5ss, alternative 5′ splice site; me, mutually exclusive exon; complex, multiexonic rearrangement; 3ss, alternative 3′ splice site; ir, intron retention) and (b) involvement in apoptosis and/or cell proliferation regulation, as well as information on the category of genes knocked down, is color coded.

(B) Boxplot representation of the mean absolute Z score of AS changes induced by the knockdown of particular classes of factors, including core and noncore SFs, other RNA processing factors, and chromatin remodeling factors. Median and spread (interquartile range) of AS changes of mock siRNA conditions are represented as a green line and box, respectively.

(C) As in (B) for spliceosomal factors that assemble early and stay as detectable components through the spliceosome cycle; factors that are present only transiently at early, mid, or late stages of assembly; and components of the exon junction complex (EJC).

See also Figure S3, Table S2, Table S3, and Table S4.

when knockdown of a subset of factors was carried out for 48 hr (Figure S2A), arguing against AS changes being the consequence of secondary effects of SF knockdown in these cases (see also below the discussion on IK/SMU1).

### Pervasive and Distinct Effects of Core Spliceosome Components on Alternative Splicing Regulation

The output of the screening process is summarized in the heatmap of Figure 2A. Three main conclusions can be derived from these results. First, a large fraction of the SF knockdowns have noticeable but distinct effects on AS, indicating that knockdown of many components of the spliceosome causes switches in splice site selection, rather than a generalized inhibition of splicing of every intron. Second, a substantial fraction of AS changes correspond to higher levels of alternative exon inclusion, again arguing against simple effects of decreasing splicing activity upon knockdown of general SFs, which would typically favor skipping of alternative exons that often harbor weaker splice sites. These observations suggest extensive versatility of the effects of modulating the levels of SFs on splice site selection. Third, despite the diversity of the effects of SF knockdowns on multiple ASEs, similarities can also be drawn (Figure 2A). For example, a number of ASEs relevant for the control of programmed cell death (e.g., Fas, APAF1, CASP9, or MCL1) cluster together and are similarly regulated by a common set of core spliceosome components. This observation suggests coordinated regulation of apoptosis by core factors, although the functional effects of these AS changes appear to be complex (Figure S3A). Another example is two distinct ASEs in the fibronectin gene (cassette exons EDA and EDB, of paramount importance to control aspects of development and cancer progression; Muro et al., 2003) which cluster close together, suggesting common mechanisms of regulation of the ASEs in this locus.

To explore the effects of different classes of SFs on AS regulation, we first grouped them in four categories: genes coding for core spliceosome components, noncore SFs/regulators, RNA-processing factors not directly related to splicing, and chromatin-related factors. Core SFs display the greater spread and average magnitude of effects, followed by noncore SFs (Figure 2B). Factors related with chromatin structure and remodeling showed a narrower range of milder effects (Figure 2B), although their values were clearly above the range of changes observed in mock knockdown samples. Of interest, knockdown of core factors leads to stronger exon skipping effects, while noncore SFs and other RNA processing and chromatin-related factors show more balanced effects (Figures S3B and S3C).

To further dissect the effects of core components, we subdivided them into four categories depending on the timing and duration of their recruitment to splicing complexes during spliceosome assembly (Wahl et al., 2009): (1) persistent components (e.g., 17S U2 snRNPs, Sm proteins) that enter the assembly process before or during complex A formation and remain until completion of the reaction, (2) transient early components that join the reaction prior to B complex activation but are scarce at later stages of the process, (3) transient middle components joining during B-act complex formation, and (4) transient late components that are only present during or after C complex formation. Our results indicate that knockdowns of factors that persist during spliceosome assembly cause AS changes of higher magnitude than those caused by knockdowns of transient factors involved in early spliceosomal complexes, which in turn are stronger and more dispersed than those caused by factors involved in middle complexes and in complex C formation or catalytic activation (Figure 2C). As observed for core factors in general, knockdown of persistent factors tends to favor skipping, while the effects of knockdown of more transient factors are more split between inclusion and skipping (Figures S3B and S3C).

Interestingly, knockdowns of components of the exon junction complex (EJC) display a range of effects that resembles those of transient early or mid splicing complexes (Figure 2C). These results are in line with recent reports documenting effects of EJC components in AS, in addition to their standard function in post-splicing processes (Ashton-Beaucage et al., 2010).

To test whether the knockdown of core factors causes a general inhibition of splicing, we measured the levels of introns (both in constitutive and alternatively spliced regions) relative to exons in four genes, upon the knockdown of four SFs (BRR2, SF3B1, SNRPG, and SLU7). The results shown in Figure 3 reveal a complex picture in which retention of certain introns is clearly increased and retention of other introns is very mild, and there is even one case (the two introns flanking the fibronectin EDI alternative exon) in which the low levels of intron detected under control conditions are reduced upon knockdown of core SFs (notably, this ASE typically displays a distinct pattern of AS upon knockdown of core SFs, Figure 2A). Therefore, rather than a general uniform inhibition of splicing, reduction in the levels of core SFs induces differential effects in different introns, a concept reminiscent of the differential effects observed for antitumor drugs targeting core components like SF3B1 (Bonnal et al., 2012).

## Reconstruction of a Functional Network of Splicing Factors

To systematically and accurately map the functional relationships of SFs on AS regulation, we quantified the similarity between AS perturbation profiles for every pair of factors in our screen using a robust correlation estimate for the effects of the factors' knockdowns across the ASEs. This measure captures the congruence between the shapes of perturbation profiles, while it does not take into account proportional differences in the magnitude of the fluctuations. Crucially, it discriminates between biological and technical outliers and is resistant to distorting effects of the latter (Supplemental Experimental Procedures).

We next employed glasso, a regularization-based algorithm for graphical model selection (Friedman et al., 2008; Supplemental Experimental Procedures), to reconstruct a network from these correlation estimates (Figures 4A and S4). Glasso seeks a parsimonious network model for the observed correlations. This is achieved by specifying a regularization parameter, which serves as a penalty that can be tuned in order to control the number of inferred connections (network sparsity) and therefore the false discovery rate (FDR) of the final model. For a given regularization parameter, both the number of inferred connections and the FDR are a decreasing function of the number of ASEs assayed (Figure 4B). Random sampling with different subsets of the real or reshuffled versions of the data indicates that network reconstruction converges to ~500 connections with a FDR <5% near 35 events (Figure 4B), implying that analysis of additional ASEs in the screening would have only small effects on network sparsity and accuracy.

The complete network of functional interactions among the screened factors is shown in Figure 4A. It is comprised of 196 nodes representing screened factors and 541 connections (average degree 5.5), of which 518 correspond to positive and 23 to negative functional associations. The topology of the resulting network has several key features. First, two classes of factors are easily distinguishable. The first class encompasses factors that form densely connected clusters comprised of core spliceosomal components (Figure 4A). Factors physically linked within U2 snRNP and factors functionally related with U2 snRNP activity, and with the transition from complex A to B, form a tight cluster (orange nodes in the inset of Figure 4A). An adjacent but distinct highly linked cluster corresponds mainly to factors physically associated within U5 snRNP or the U4/5/6

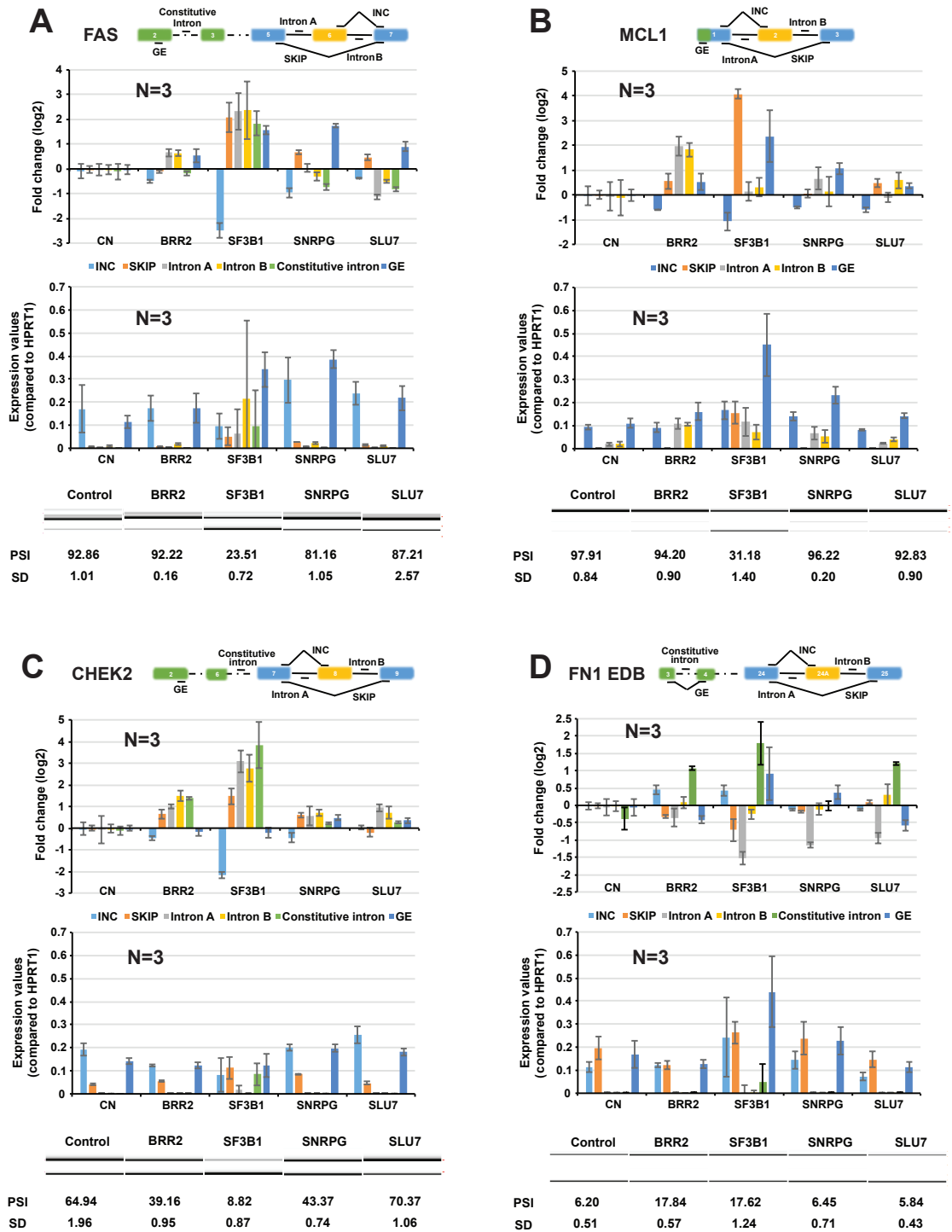

**Figure 3. Variety of Effects of Knockdown of Core SFs on Intron Retention**

(A–D) Real-time quantification of AS, intron retention, and gene expression for four genes included in the splicing network (FAS, A; MCL1, B; CHEK2, C; and FN1, D) upon knockdown of four core spliceosomal components (BRR2, SF3B1, SNRPG, and SLU7). In each of the panels, the graphs represent the following: top, scheme of genomic locations, AS patterns, and amplicons used; middle top, quantification of fold changes in expression of the different isoforms or introns (indicated by the color code) compared to mock siRNA conditions; middle bottom, relative levels of the different isoforms or introns (indicated by the color code) compared to HPRT1 housekeeping gene mRNAs; and bottom, AS changes measured by HTCE. PSI values are indicated. For all assays, values represent the mean, and error bars the standard deviation of three independent biological replicas.

tri-snRNP (blue nodes in Figure 4A, inset). Factors in these clusters and their mutual links encompass 23% of the network nodes but 63% of the total inferred functional associations (average degree 14.5). A second category includes factors that form a periphery of lower connectivity that occasionally projects to the core. This category includes many of the classical splicing regulators (e.g., SR proteins, hnRNPs) as well as several chromatin factors. An intermediate category—more densely interconnected but not reaching the density of interactions of the central modules—includes multiple members of the RNA-dependent DEAD/X box helicase family. While the number of connections (degree) is significantly higher for core spliceosomal factors in general (Figure 4C), this is especially clear for factors that represent persistent components of splicing complexes along the assembly pathway (Figures 4D and 4E). Persistent factors are particularly well connected to each other (50% of the possible functional connections are actually detected, Figure 4E). Factors that belong to consecutive complexes are significantly more linked to each other than factors that belong to early and late complexes, and factors that transiently assemble toward the final stages of spliceosome assembly display the least number of connections (Figures 4D and 4E).

Collectively, these results indicate that core SFs, particularly those that persist during assembly, display highly related functions in splice site recognition and fluctuations in their levels or activities have related consequences on the modulation of splice site choice. Of the observed connections, 20% can be attributed to known physical interactions and 82% of these are among core components (Figure 4A). Examples include tight links between the two subunits of U2AF, between the interacting partners PLRG1 and CDC5L or IK and SMU1. We estimate that about 50% of the functional links detected in our network could be predicted by previous knowledge of the composition or function of splicing complexes.

Of relevance, a substantial number of the functional associations closely recapitulate the composition and even—to some extent—the topology of known spliceosomal complexes (Figure S4; see also below and Discussion). These observations suggest that differences in the sensitivity of ASEs to SF perturbations reflect the specific role of these factors in the splicing process. Conversely, these results warrant the use of splicing perturbation profiles as proxies for inferring physical or functional links between factors in the splicing process.

## Generality of the Functional Interactions in the AS Regulation Network

The above network was derived from variations in 35 ASEs selected because of their relevance for cell proliferation and apoptosis, but they represent a heterogeneous collection of AS types and, possibly, regulatory mechanisms. To identify functional connections persistent across AS types, we undertook a subsampling approach. We sampled subsets of 17 from the original 35 ASEs and iteratively reconstructed the network of functional associations from each subset (see Supplementary methods). Following 10,000 iterations, we asked how many functional connections were recovered in at least 90% of the sampled networks. The retrieved set of connections are shown in Figure 5A and they can be considered to represent a "basal" or indispensable splicing circuitry that captures close similarities in function between SFs regardless of the subset of splicing substrates considered. The majority of these connections correspond to links between core components, with modules corresponding to major spliceosomal complexes. For example, the majority of components of U2 snRNP constitute the most prominent of the core modules (Figure 5A). Strikingly, the module is topologically subdivided in two submodules, the first corresponding to components of the SF3a and SF3b complexes and the second corresponding to proteins in the Sm complex. The connectivity between U2 snRNP components remarkably recapitulates known topological features of U2 snRNP organization, including the assembly of SF3a/b complexes in stem loops I and II and the assembly of the Sm ring in a different region of the U2 snRNA (Behrens et al., 1993) (Figures 5A and S4). That the connectivity between components derived from assessing the effects on AS of depletion of these factors correlates with topological features of the organization of the snRNP further argues that the functions in splice site selection are tightly linked to the structure and function of this particle, highlighting the potential resolution of our approach for identifying meaningful structural and mechanistic links.

A second prominent "core" module corresponds to factors known to play roles in conformational changes previous to/ concomitant with catalysis in spliceosomes from yeast to human, including PRP8, PRP31 or U5-200. Once again some of the topological features of the module are compatible with physical associations between these factors (e.g., PRP8 with PRP31 or U5-116K or PRP31 with C20ORF14). That knockdown of these late-acting factors, which coordinate the final steps of the splicing process (Bottner et al., 2005; Häcker et al., 2008; Wahl et al., 2009), causes coherent changes in splice site selection strongly argues that the complex process of spliceosome assembly can be modulated at late steps, or even at the time of catalysis, to affect splice site choice. Other links in the "core" network recapitulate physical interactions, including the aforementioned modules involving the two subunits of the 3' splice site-recognizing factor U2AF, the interacting partners PLRG1 and CDC5L and IK and SMU1.

To further evaluate the generality of our approach, we tested whether the functional concordance derived from our network analysis and from the iterative selection of a "core" network could be recapitulated in a different cell context and genome-wide. We focused our attention in the strong functional association between IK and SMU1, two factors that have been described as associated with complex B and its transition to Bact (Bessonov et al., 2008) (Figure 5A). Analysis of the effects of knockdown of these factors in the same sets of ASEs in HEK293 cells revealed a similar set of associations, despite the fact that individual events display different splicing ratios upon IK/SMU1 knockdown in this cell line compared to HeLa (Figures 5B and 5C). To explore the validity of this functional association in a larger set of splicing events, RNA was isolated in biological triplicates from HeLa cells transfected with siRNAs against these factors, and changes in AS were assessed using genome-wide splicing-sensitive microarrays. The results revealed a striking overlap between the effects of IK and SMU1 knockdown, both on gene expression changes and on

**A**

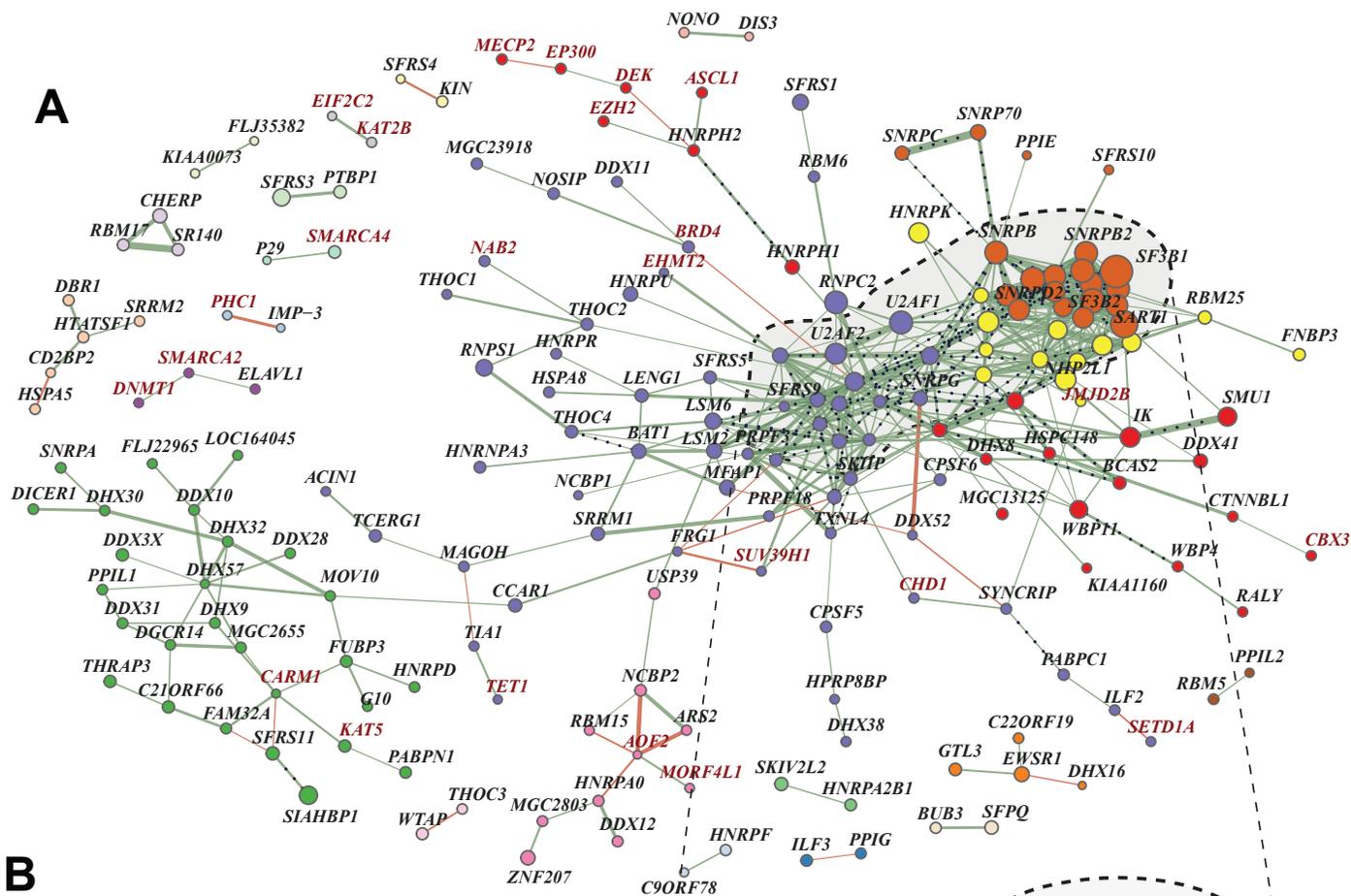

**B**

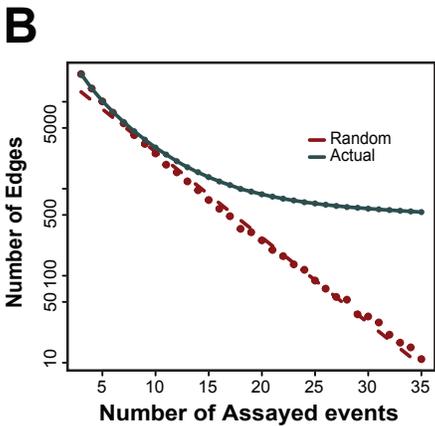

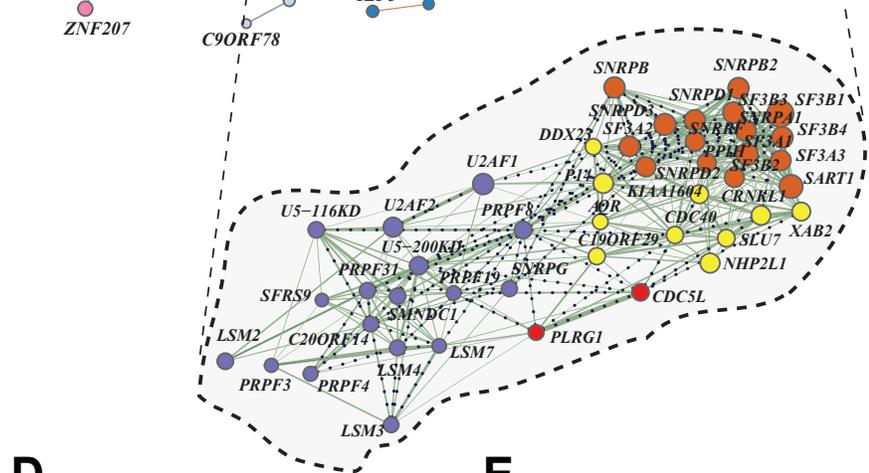

**C**

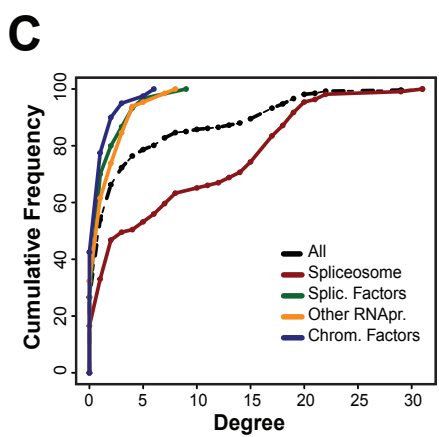

**D**

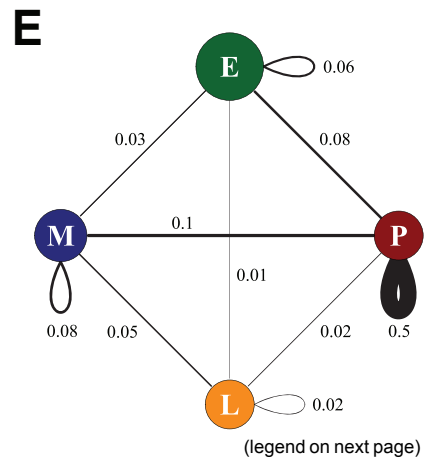

**E**

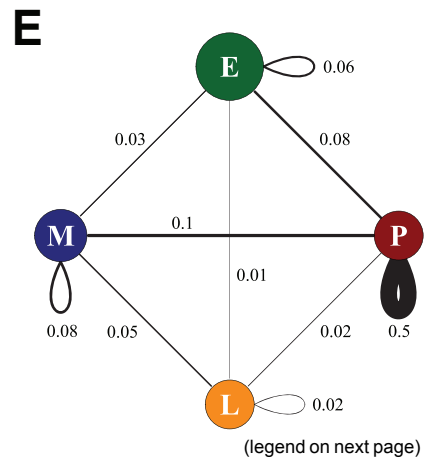

AS changes (61% overlap, p < 1E-20) covering a wide range of gene expression levels, fold differences in isoform ratios and AS types (Figures 5D–5F, S5B, and S5C). The overlap was much more limited with RBM6, a splicing regulator (Bechara et al., 2013) located elsewhere in the network. These results validate the strong functional similarities between IK and SMU1 detected in our analysis. The molecular basis for this link could be explained by the depletion of each protein (but not their mRNA) upon knockdown of the other factor, consistent with formation of a heterodimer of the two proteins and destabilization of one partner after depletion of the other (Figure S5A). Notably, the best correlation between the effects of these factors was between IK knockdown for 48 hr and SMU1 knockdown for 72 hr, perhaps reflecting a stronger functional effect of IK on AS regulation (Figures S2B and S2D).

Gene ontology analysis revealed a clear common enrichment of AS changes in genes involved in cell death and survival (Figure S5D), suggesting that IK/SMU1 could play a role in the control of programmed cell death through AS. To evaluate this possibility, we tested the effects of knockdown of these factors, individually or combined, on cell proliferation and apoptotic assays. The results indicated that their depletion activated cleavage of PARP, substantially reduced cell growth, and increased the fraction of cells undergoing apoptosis (Figures S5E–S5G). We conclude that functional links revealed by the network can be used to infer common mechanisms of splicing regulation and provide insights into the biological framework of the regulatory circuits involved.

## Ancillary Functional Network Interactions Reveal Alternative Mechanisms of AS Regulation

The iterative generation of networks from subsets of ASEs explained above could, in addition to identifying general functional interactions, be used to capture functional links that are specific to particular classes of events characterized by distinct regulatory mechanisms. To explore this possibility, we sought the set of functional connections that are robustly recovered in a significant fraction of ASEs subsets but are absent in others (see Supplemental Experimental Procedures). Figure 6A shows the set of such identified interactions, which therefore represents specific functional links that only emerge when particular subsets of ASEs are considered. The inset represents graphically the results of principal component analysis (PCA) showing the variability in the strength of these interactions depending on the presence or absence of different ASEs across the subsamples. The most discriminatory set of ASEs for separating these interactions is also shown (see also Figure S6). While this analysis may be constrained by the limited number of events analyzed and higher FDR due to multiple testing, it can provide a valuable tool for discovering regulatory mechanisms underlying specific classes of ASEs as illustrated by the example below.

A synergistic link was identified between hnRNP C and U2AF1 (the 35 kDa subunit of U2AF, which recognizes the AG dinucleotide at 3' splice sites) for a subset of ASEs (Figures 6A and S6A). We explored a possible relationship between the effects of knocking down these factors and the presence of consensus hnRNP C binding sites (characterized by stretches of uridine residues) or 3' splice site-like motifs in the vicinity of the regulated exons. The results revealed a significant correlation between the two factors (0.795), associated with the presence of composite elements comprised of putative hnRNP C binding sites within sequences conforming to a 3' splice site motif (see Experimental Procedures) upstream and/or downstream of the regulated exons (Figures 6B and S6B). No such correlation was observed for ASEs lacking these sequence features (Figures 6C and S6C). These observations are in line with results from a recent report revealing that hnRNP C prevents exonization of Alu elements by competing the binding of U2AF to Alu sequences (Zarnack et al., 2013). Our results extend these findings by capturing functional relationships between hnRNP C and U2AF in ASEs that harbor binding sites for these factors in the vicinity of the regulated splice sites, both inside of Alu elements or independent of them (Figures 6B and 6D). Our data are compatible with a model in which uridine-rich and 3' splice site-like sequences sequester U2AF away from the regulated splice sites, causing alternative exon skipping, while hnRNP C displaces U2AF away from these decoy sites, facilitating exon inclusion. In this model, depletion of U2AF and hnRNP C are expected to display the same effects in these ASEs, as observed in our data (Figures 6B and 6D). In contrast, in ASEs in which hnRNP C binding sites are located within the polypyrimidine tract of the 3' splice site of the regulated exon, the two factors display antagonistic effects (Figures 6C and 6D), as expected from direct competition between them at a functional splice site. This example illustrates the potential of the network approach to identify molecular mechanisms of regulation on the basis of specific sequence features and the interplay between cognate factors.

---

**Figure 4. Functional Splicing Regulatory Network**

(A) Graphical representation of the reconstructed splicing network. Nodes (circles) correspond to individual factors and edges (lines) to inferred functional associations. Positive or negative functional correlations are represented by green or red edges, respectively. Edge thickness signifies the strength of the functional interaction, while node size is proportional to the overall impact (median Z score) of a given knockdown in the regulation of AS. Node coloring depicts the network's natural separation in coherent modules (see Supplemental Experimental Procedures). Known physical interactions as reported in the STRING database (Franceschini et al., 2013) are represented in black dotted lines between factors. The inset is an expanded view of factors in the U2 and U4/5/6 snRNP complexes.

(B) Number of recovered network edges using actual or randomized data sets (red and green points respectively) as a function of the number of ASEs used. Lines represent the best fit curves for the points.

(C and D) Network degree distribution. The plots represent the cumulative frequency of the number of connections (degree) for different classes of factors (as in Figures 2B and 2C).

(E) Functional crosstalk between different categories of spliceosome components. Values represent the fraction of observed functional connections out of the total possible connections among factors that belong to the different categories: persistent, P; transient early, E; transient mid, M; or transient late, L factors. See also Figure S4E and Table S5.

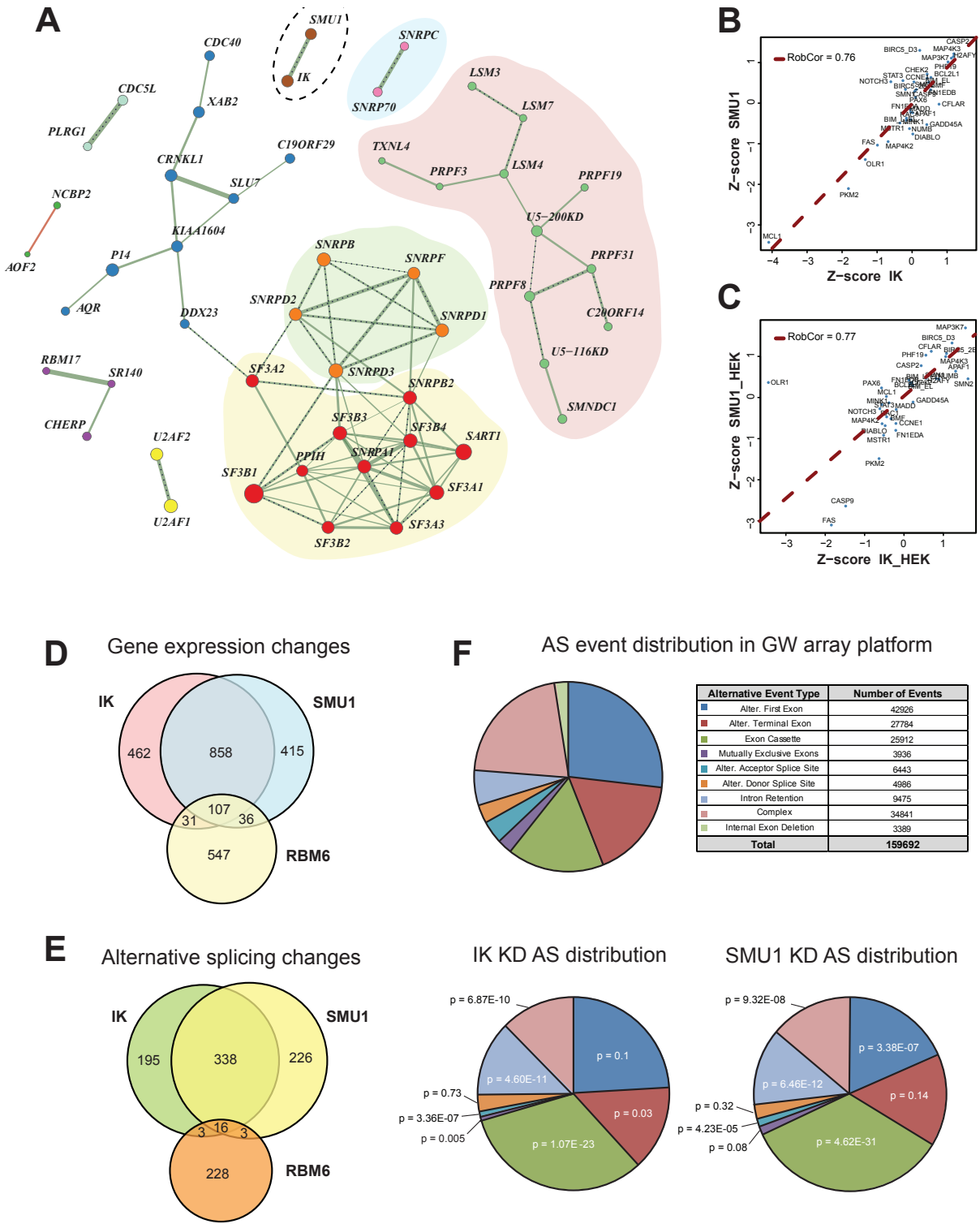

**Figure 5. Core Network Involved in AS Regulation**

(A) Graphical representation of the functional connections present in at least 90% of 10,000 networks generated by iterative selection of subsets of 17 out of the 35 ASEs used to generate the complete network. Network attributes represented as in Figure 4A. Known spliceosome complexes are highlighted by shadowed areas (U1 snRNP, blue; U2 snRNP, yellow; SM proteins, green; and tri-snRNP proteins, red).

(B and C) Consistency of inferred functional interactions in different cell lines. Knockdown of IK or SMU1 was carried out in parallel in HeLa (B) or HEK293 cells (C), and changes for the 35 ASEs were analyzed by RT-PCR and HTCE. Robust correlation estimates and regression for the AS changes observed in IK versus SMU1 knockdowns are shown for each of the cell lines.

(D) Venn diagrams of the overlap between the number of gene expression changes upon IK, SMU1, or RBM6 knockdowns.

(legend continued on next page)

## Mapping Drug Targets within the Spliceosome

Another application of our network analysis is to identify possible targets within the splicing machinery of physiological or external (e.g., pharmacological) perturbations that produce changes in AS that can be measured using the same experimental setting. The similarity between profiles of AS changes induced by such perturbations and the knockdown of particular factors can help to uncover SFs that mediate their effects. To evaluate the potential of this approach, cells were treated with drugs known to affect AS decisions and the patterns of changes in the 35 ASEs induced by these treatments were assessed by the same robotized procedure for RNA isolation and RT-PCR analysis used to locate their position within the network.

The results indicate that structurally similar drugs like Spliceostatin A (SSA) and Meayamycin cause AS changes that closely resemble the effects of knocking down components of U2 snRNP (Figure 7A), including close links between these drugs and SF3B1, a known physical target of SSA (Kaida et al., 2007; Hasegawa et al., 2011) previously implicated in mediating the effects of the drug through alterations in the AS of cell cycle genes (Corrionero et al., 2011). Of interest, changes in AS of MCL1 (leading to the production of the proapoptotic mRNA isoform) appear as prominent effects of both drugs (Figure 7B), confirming and extending recent observations obtained using Meayamycin (Gao et al., 2013) and suggesting that this ASE can play a key general role in mediating the antiproliferative effects of drugs that target core splicing components like SF3B1 (Bonnal et al., 2012). A strong link is also captured between each of the drugs and PPIH (Figure 7A), a peptidyl-prolyl cis-trans isomerase which has been found to be associated with U4/5/6 tri-snRNP (Horowitz et al., 1997; Teigelkamp et al., 1998) but had not been implicated directly in 3′ splice site regulation. Indeed, the network indicates a close relationship between PPIH and multiple U2 snRNP components, particularly in the SF3a and SF3b complexes, further arguing that PPIH plays a role in 3′ splice site definition closely linked to branchpoint recognition by U2 snRNP. Treatment of the cells with the Clk (Cdc2-like) kinase family inhibitor TG003, which is also known to modulate AS (Muraki et al., 2004) and causes changes in ASEs analyzed in our network (Figure 7B), led to links with a completely different set of SFs (Figure 7A), attesting to the specificity of the results. These data confirm the potential of the network analysis to identify bona fide SF targets of physiological or pharmacological perturbations and further our understanding of the the underlying molecular mechanisms of regulation.

## DISCUSSION

The combined experimental and computational approach described in this manuscript provides a rich source of information and a powerful toolset for studying the splicing process and its regulation. Similar approaches could be useful to systematically capture information about other aspects of transcriptional or posttranscriptional gene regulation. The data provide comprehensive information about SFs (and some additional chromatin factors) relevant to understand the regulation of 35 ASEs important for cell proliferation and apoptosis. Second, the network analysis reveals functional links between SFs, which in some cases are based upon physical interactions. The reconstruction of the known topology of some complexes (e.g., U2 snRNP) by the network suggests that the approach captures important aspects of the operation of these particles and therefore has the potential to provide novel insights into the composition and intricate workings of multiple Spliceosomal subcomplexes. Third, the network can serve as a resource for exploring mechanisms of AS regulation induced by perturbations of the system, including genetic manipulations, internal or external stimuli, or exposure to drugs. As an example, the network was used in the accompanying manuscript by Tejedor et al. (2014) to identify a link between AS changes induced by variations in intracellular iron and the function of SRSF7, which can be explained in terms of iron-induced changes in the RNA binding and splicing regulatory activity of this zinc-knuckle-containing SR protein. Other examples include (1) the identification of the extensive regulatory overlap between the interacting proteins IK/RED and SMU1, relevant for the control of programmed cell death; (2) evidence for a general role of uridine-rich hnRNP C binding sequences (some associated with transposable elements; Zarnack et al., 2013) in the regulation of nearby alternative exons via antagonism with U2AF; and (3) delineation of U2 snRNP components and other factors, including PPIH, that are functionally linked to the effects of antitumor drugs on AS.

Our data reveal changes of AS regulation mediated either by core splicing components or by classical regulatory factors like proteins of the SR and hnRNP families. While the central core of functional links is likely to operate globally, alternative configurations (particularly in the periphery of the network) are likely to emerge if similar approaches are applied to other cell types, genes, or biological contexts. We report considerable versatility in the effects of core spliceosome factors on alternative splicing. Precedents exist for a role of core SFs on the regulation of splicing efficiency or splice site selection in yeast, Drosophila, and mammalian cells (Clark et al., 2002; Park et al., 2004; Pleiss et al., 2007; Saltzman et al., 2011). Despite their assumed general roles in the splicing process, depletion or mutation of particular core factors led to differential splicing effects in these studies, and at least in some cases the differential effects could be attributed to particular features of the regulated pre-mRNAs. For example, Clark et al. (2002) found that the core components Prp17p and Prp18p are dispensable for splicing of yeast introns with short branchpoint to 3′ splice site distances. Pleiss et al. (2007) found that introns of yeast ribosome protein genes are particularly sensitive to mutation of various core factors, including two DEAD/X box family of RNA helicases (PRP2 and PRP5) involved in spliceosome conformational transitions and PRP8, a highly conserved protein involved in catalytic activation

(E) Venn diagrams of the overlap between the number of AS changes upon knockdown of IK, SMU1, or RBM6.
(F) Distribution of ASEs in the Affymetrix array platform (upper panel) and distribution of the AS changes observed upon IK (lower left) or SMU1 (lower right) knockdown. p values correspond to the difference between the distribution of AS changes upon knockdown and the distribution of ASEs in the array. See also Figure S5.

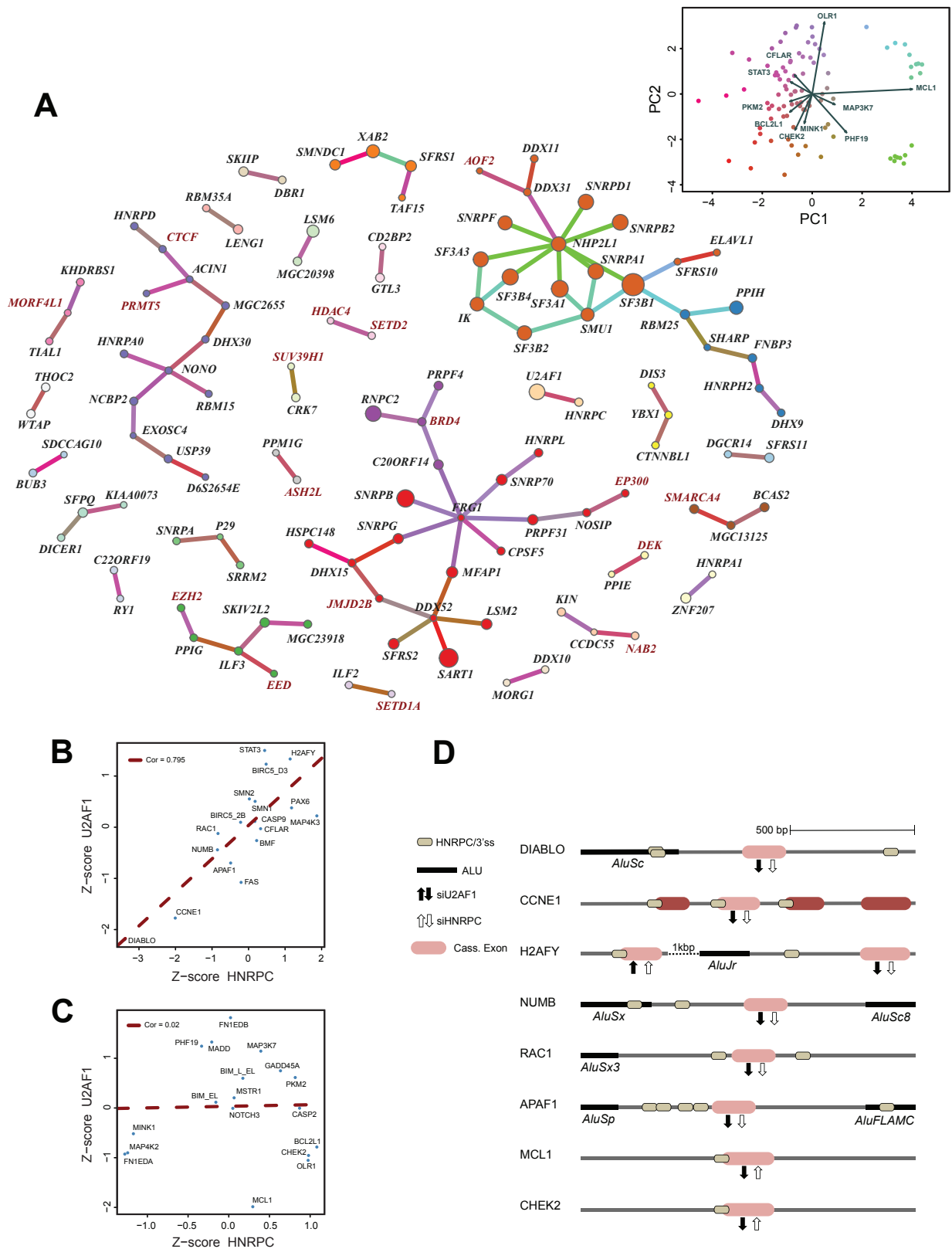

**Figure 6. Variable Functional Interactions Extracted from Networks Generated from Subsets of ASEs**

(A) Network of ancillary functional connections characteristic of subsets of ASEs. Lines link SFs that display functional similarities in their effects on subsets of ASEs. Line colors indicate the positions of the interactions in the PCA shown in the inset, which captures the different relative contributions of the ASEs in defining these connections (see also Figure S6 and Experimental Procedures). For clarity, only the top ten discriminatory ASEs are shown.

(B and C) Correlation values and regression for the effects of U2AF1 versus HNRPC knockdowns for ASEs in the presence (B) or absence (C) of upstream composite HNRPC binding/3' splice site-like elements.

(legend continued on next page)

(Collins and Guthrie, 1999; Mozaffari-Jovin et al., 2012). Distinct effects of the mutations on different ribosomal genes were also observed, arguing again for specific effects on splicing of certain genes. DEAD/H box proteins as well as components of U1, U2, and U4/6 snRNP were found to modulate particular ASEs in an RNAi screen of RNA binding proteins in Drosophila cells (Park et al., 2004). In contrast with the standard function of classical splicing regulators acting through cognate binding sites specific to certain target RNAs, core factors could both carry out general, essential functions for intron removal and in addition display regulatory potential if their levels become limiting for the function of complexes, leading to differences in the efficiency or kinetics of assembly on alternative sites. This could be the case for the knockdown of SmB/B[0], a component of the Sm complex present in most spliceosomal snRNPs, which leads to autoregulation of its own pre-mRNA as well as to effects on hundreds of other ASEs, particularly in genes encoding RNA processing factors (Saltzman et al., 2011).

The results of our genome-wide siRNA screen for regulators of Fas AS (Tejedor et al., 2014) provide unbiased evidence that an extensive number of core SFs have the potential to contribute to the regulation of splice site selection. This is, in fact, the most populated category of screen hits. Similarly, the results of our network highlight coherent effects on alternative splice site choice of multiple core components, revealing also that the extent of their effects on alternative splice site selection can be largely attributed to the duration and order of their recruitment in the splicing reaction.

Remarkably, effects on splice site selection are associated with depletion not only of complexes involved in early splice site recognition but also of factors involved in complex B formation or even in catalytic activation of fully assembled spliceosomes. While particular examples of AS regulation at the time of the transition from complex A to B and even at later steps had been reported (House and Lynch, 2006; Bonnal et al., 2008; Lallena et al., 2002), our results indicate that the implication of late factors is quite general. Such links, e.g., those involving PRP8 and other interacting factors closely positioned near the reacting chemical groups at the time of catalysis, like PRP31 or U5 200KD, are actually part of the persistent functional interactions that emerge from the analysis of any subsample of ASEs analyzed, highlighting their general regulatory potential in splice site selection. How can factors involved in late steps of the splicing process influence splice site choices? It is conceivable that limiting amounts of late spliceosome components would favor splicing of alternative splice site pairs harboring early complexes that can more efficiently recruit late factors, or influence the kinetics of conformational changes required for catalysis. In this context, the regulatory plasticity revealed by our network analysis may be contributed, at least in part, by the emerging realization that a substantial number of SFs contain disordered regions when analyzed in isolation (Korneta and Bujnicki, 2012; Chen and Moore, 2014). Such regions may be flexible to adopt different conformations in the presence of other spliceosomal components, allowing alternative routes for spliceosome assembly on different introns, with some pathways being more sensitive than others to the depletion of a general factor.

Kinetic effects on the assembly and/or engagement of splice sites to undergo catalysis would be particularly effective if spliceosome assembly is not an irreversible process. Results of single molecule analysis are indeed compatible with this concept (Tseng and Cheng, 2008; Abelson et al., 2010; Hoskins et al., 2011; Hoskins and Moore, 2012; Shcherbakova et al., 2013), thus opening the extraordinary complexity of conformational transitions and dynamic compositional changes of the spliceosome as possible targets for regulation. Interestingly, while depletion of early factors tends to favor exon skipping, depletion of late factors causes a similar number of exon inclusion and skipping effects, suggesting high plasticity in splice site choice at this stage of the process.

Given this potential, it is conceivable that modulation of the relative concentration of core components can function as a physiological mechanism for splicing regulation, for example during development and cell differentiation. Indeed, variations in the relative levels of core spliceosomal components have been reported (Wong et al., 2013). Furthermore, recent reports revealed the high incidence of mutations in core SFs in cancer. For example, the gene encoding SF3B1, a component of U2 snRNP involved in branchpoint recognition, has been described as among the most highly mutated in myelodysplastic syndromes (Yoshida et al., 2011), chronic lymphocytic leukemia (Quesada et al., 2012), and other cancers (reviewed in Bonnal et al., 2012), correlating with different disease outcomes. Remarkably, SF3B1 is the physical target of antitumor drugs like Spliceostatin A and—likely—Meayamycin, as captured by our study. These observations are again consistent with the idea that depletion, mutation, or drug-mediated inactivation of core SFs may not simply cause a collapse of the splicing process—at least under conditions of limited physical or functional depletion—but rather lead to changes in splice site selection that can contribute to physiological, pathological, or therapeutic outcomes. SF3B1 participates in the stabilization and proofreading of U2 snRNP binding to the branchpoint region (Corrionero et al., 2011), and it is therefore possible that variations in the activity of this protein result in switches between alternative 3' splice sites harboring branch sites with different strengths and/or flanking decoy binding sites. Similar concepts may apply to explain the effects of mutations in other SFs linked to disease, including, for example, mutations in PRP8 leading to retinitis pigmentosa (Pena et al., 2007).

In summary, the data and methodological approaches presented in this study provide a rich resource for understanding the function of the spliceosome and the mechanisms of AS regulation, including the identification of targets of physiological, pathological, or pharmacological perturbations within the complex splicing machinery.

---

(D) Schematic representation of the distribution of composite HNRPC binding sites/3' splice site-like sequences in representative examples of the splicing events analyzed. The relative position of Alu elements in these regions is also shown. Genomic distances are drawn to scale. The direction of the arrows indicates up or downregulation of cassette exons upon knockdown of U2AF1 (black arrows) or HNRPC (white arrows). See also Figure S6.

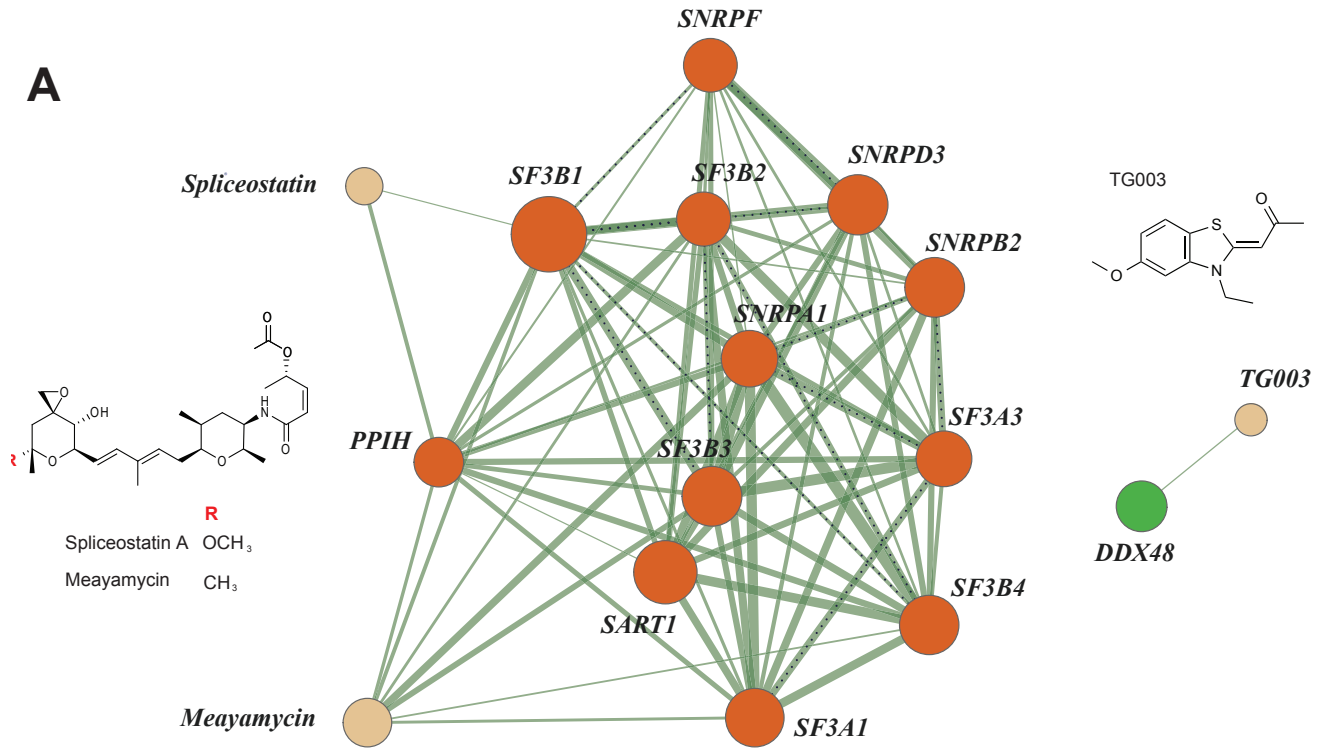

**A**

Spliceostatin

SNRPF

SF3B1 SF3B2 SNRPD3

SNRPB2

SNRPA1

PPIH SF3B3 SF3A3

SF3B4

SART1

SF3A1

Meayamycin

R

Spliceostatin A   OCH₃

Meayamycin      CH₃

TG003

TG003

DDX48

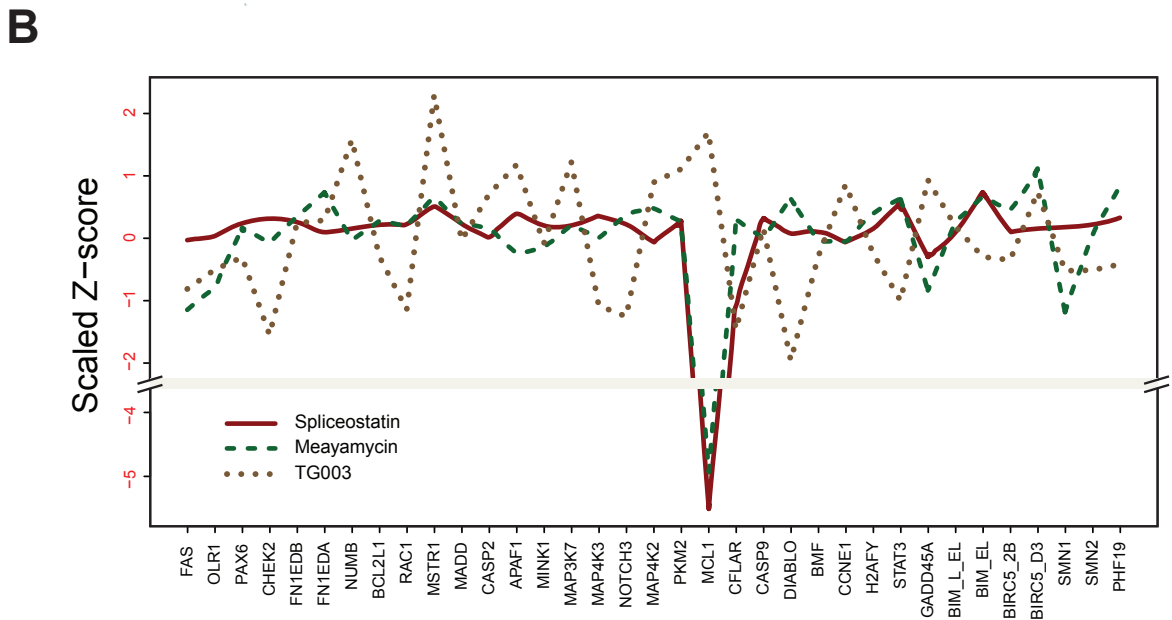

**B**

Scaled Z-score

— Spliceostatin
- - Meayamycin
····· TG003

FAS OLR1 PAX6 CHEK2 FN1EDB FN1EDA NUMB BCL2L1 RAC1 MSTR1 MADD CASP2 APAF1 MINK1 MAP3K7 MAP4K3 NOTCH3 MAP4K2 PKM2 MCL1 CFLAR CASP9 DIABLO BMF CCNE1 H2AFY STAT3 GADD45A BIM_L_EL BIM_EL BIRC5_2B BIRC5_D3 SMN1 SMN2 PHF19

## EXPERIMENTAL PROCEDURES

### siRNA Library Transfection and mRNA Extraction

HeLa cells were transfected in biological triplicates with siRNAs pools (ON TARGET plus smartpool, Dharmacon, Thermo Scientific) against 270 splicing and chromatin remodeling factors (see Table S2). Endogenous mRNAs were purified 72 hr posttransfection by using oligo dT-coated 96-well plates (mRNA catcher PLUS, Life Technologies) following the man- ufacturer's instructions. RNA samples corresponding to treatments with splicing arresting drugs were isolated with the Maxwell 16 LEV simplyRNA kit (Promega).

### RT-PCR and High-Throughput Capillary Electrophoresis

Cellular mRNAs were reverse transcribed using Superscript III Retro Tran- scriptase (Invitrogen, Life Technologies) following the manufacturer's recom- mendations. PCR reactions for every individual splicing event analyzed were carried out using forward and reverse primers in exonic sequences flanking the alternatively spliced region of interest and further reagents provided in the GOTaq DNA polymerase kit (GoTaq, Promega). Primers used in this study are listed in Table S6.

HTCE measurements for the different splicing isoforms were performed in 96-well format in a Labchip GX Caliper workstation (Caliper, Perkin Elmer) us- ing a HT DNA High Sensitivity LabChip chip (Perkin Elmer). Data values were obtained using the Labchip GX software analysis tool (version 3.0).

### Quantification of AS Changes from HTCE Measurements

Robust estimates of isoform ratios upon siRNA or pharmacological treatments were obtained using the median PSI indexes of the biological triplicates for each knockdown-AS pair. From this value $x_i$ the effect of each treatment was summarized as a robust Z score (Birmingham et al., 2009):

### Robust Sample Correlation Estimation

Robust correlation estimation is based on an iterative weighting algorithm that discriminates between technical outliers and reliable measure- ments with high leverage. The weighting for mea- surements relies on calculation of a reliability index that takes into account their cumulative influence on correlation estimates of the complete data set. Algorithmic details are provided in Supplemental Experimental Procedures.

### Network Reconstruction

The process of network reconstruction relies on the regularization-based graphical lasso (glasso) algorithm for graphical model selection (Friedman et al., 2008). The glasso process for network reconstruction was implemented using the glasso R package by Jerome Friedman, Trevor Hastie, and Rob Tibshirani (http://statweb.stanford.edu/~tibs/glasso/). A detailed exposition of the algorithmic steps for network reconstruction and analysis, glasso, and its use for graphical model selection is provided in Supplemental Experimental Procedures.

### Principal Component Analysis of Ancillary Connections

PCA for ancillary edges was performed using R-mode PCA (R function prin- comp). The input data set is the scaled 95 3 35 matrix containing the absolute average correlation for each of the 95 edges over the subsets of the 10,000 samples that included each of the 35 events. The coordinates for projecting the 95 edges are directly derived from their scores on the first two principal components. The arrows representing the top ten discriminatory events (based on the vector norm of the first two PC loadings) were drawn from the origin to the coordinate defined by the first two PC loadings scaled by a con- stant factor for display purposes.

### Splice Site Scoring, Identification of Probable HNRPC Binding Sites and Alu Elements

For the identification and scoring of 3′ ss we used custom-built position weight
matrices (PWMs) of length 21—8 intronic plus three exonic positions. The matrices were built using a set of human splice sites from constitutive, internal exons compiled from the hg19 UCSC annotation database (Karolchik et al., 2014). Background nucleotide frequencies were estimated from a set of strictly intronic regions. Threshold was set based on a FDR of 0.5/kbp of random in- tronic sequences generated using a second-order Markov chain of actual in- tronic regions.

We considered as probable HNRPC binding sites consecutive stretches of U{4} as reported in Zarnack et al. (2013). Annotation of ALU elements was based on the RepeatMasker track developed by Arian Smit (http://www. repeatmasker.org) and the Repbase library (Jurka et al., 2005).

More detailed experimental procedures are provided in Supplemental Experimental Procedures.

## AUTHOR CONTRIBUTIONS

P.P. set up the computational pipeline to generate the splicing network. J.R.T. set up and carried out the experimental analyses. L.V. contributed to the network analyses involving drugs. P.P., J.R.T., and J.V. designed the project and wrote the manuscript.

## ACKNOWLEDGMENTS

We thank many CRG colleagues, EURASNET members, and Quaid Morris for their advice, encouragement, and critical reading of the manuscript, and Drs. Koide and Yoshida for reagents. We acknowledge the excellent technical sup- port of the CRG Robotics and Genomics facilities. J.R.T. was supported by a PhD fellowship from Fondo de Investigaciones Sanitarias. Work in our lab is supported by Fundació n Botín, Consolider RNAREG, Ministerio de Economía e Innovación, and AGAUR.

# SUPPLEMENTAL INFORMATION
## ATTACHED BELOW

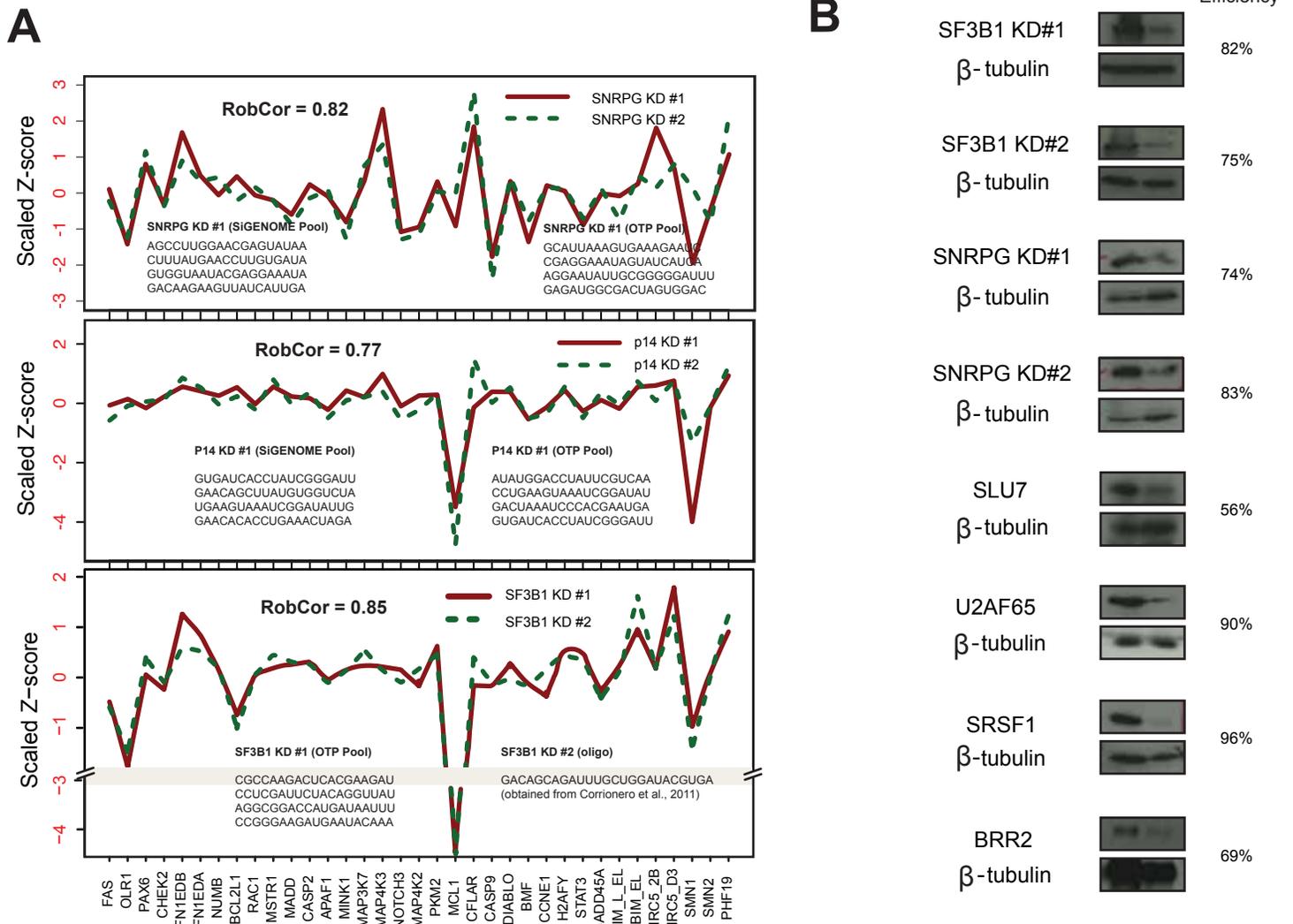
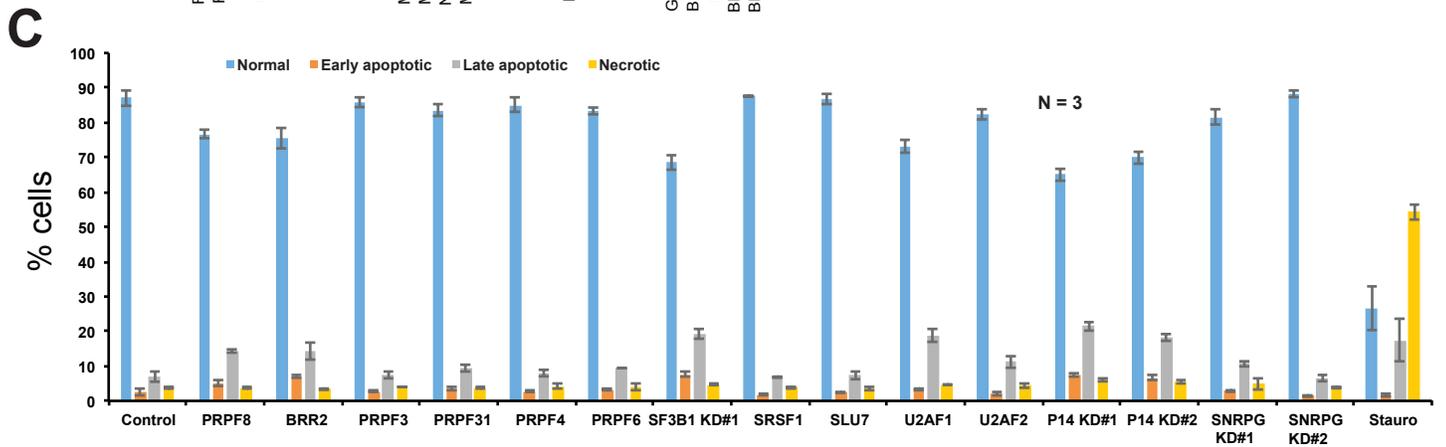
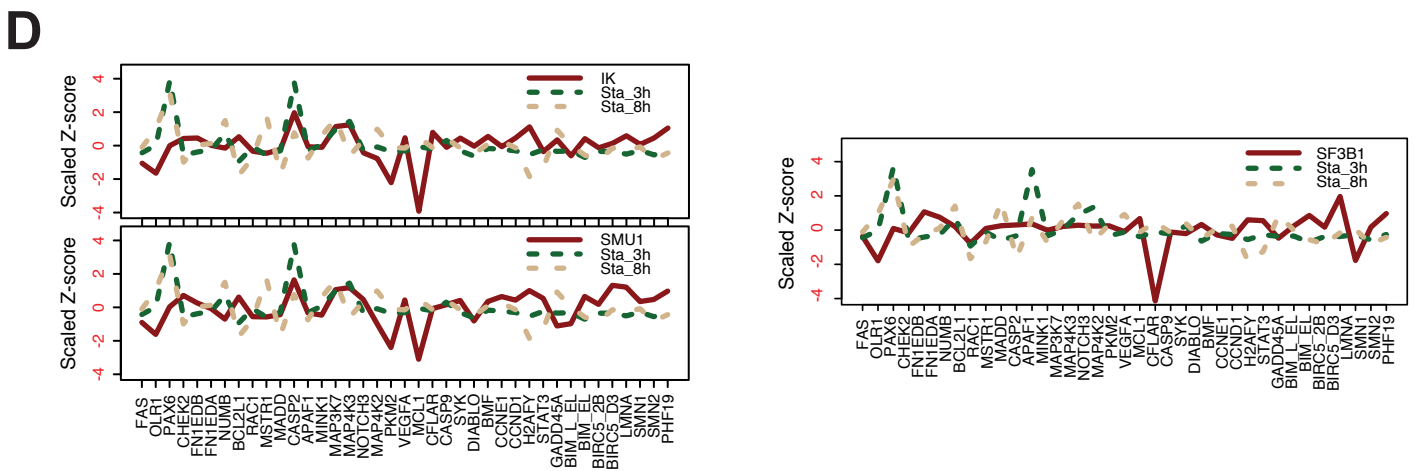

**A**

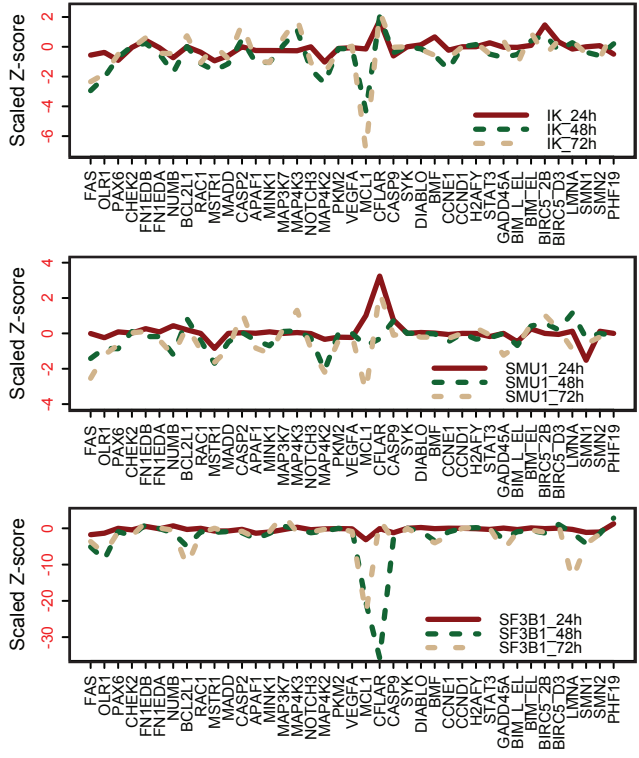

**B**

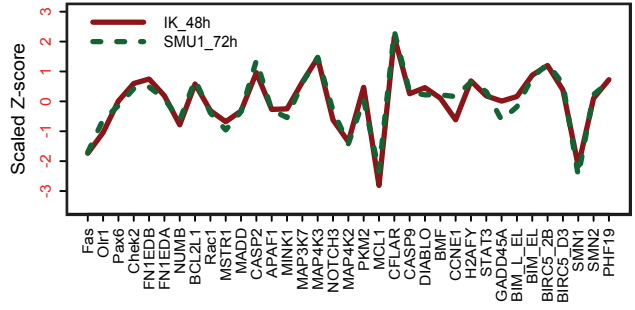

**C**

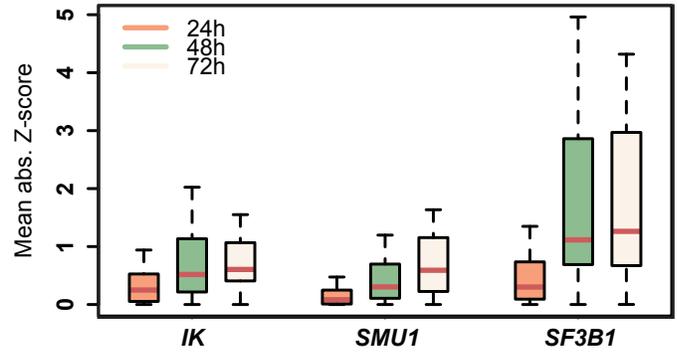

**D**

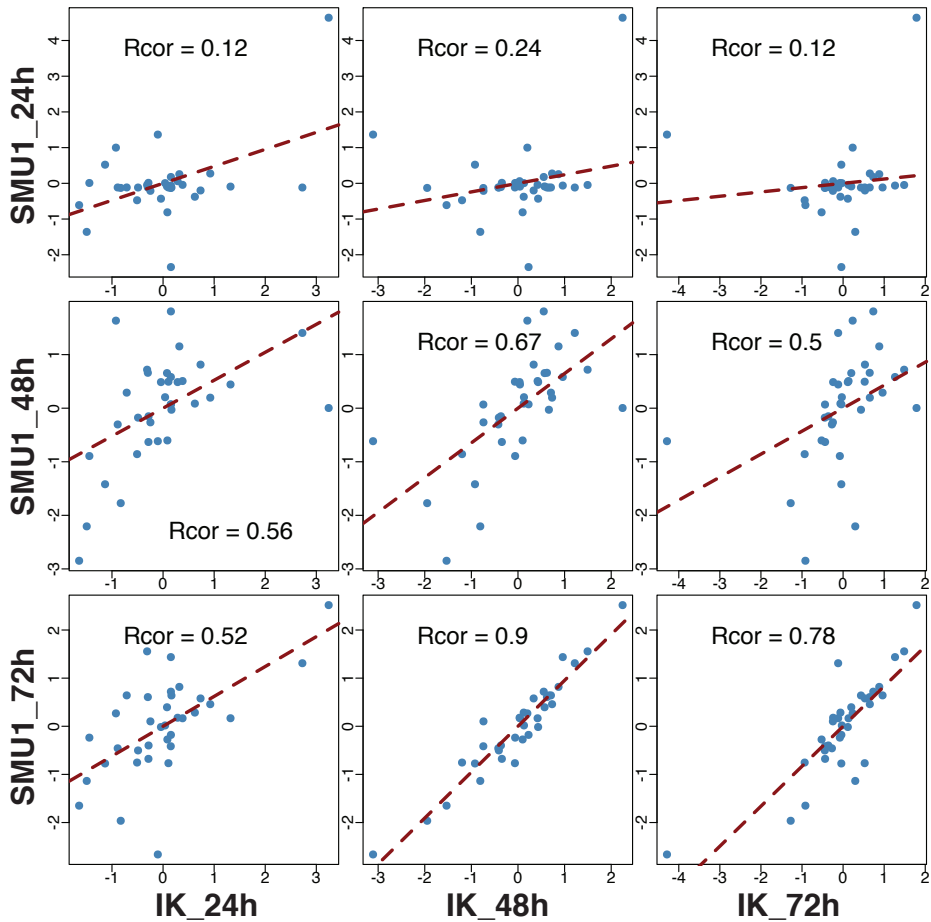

**A**

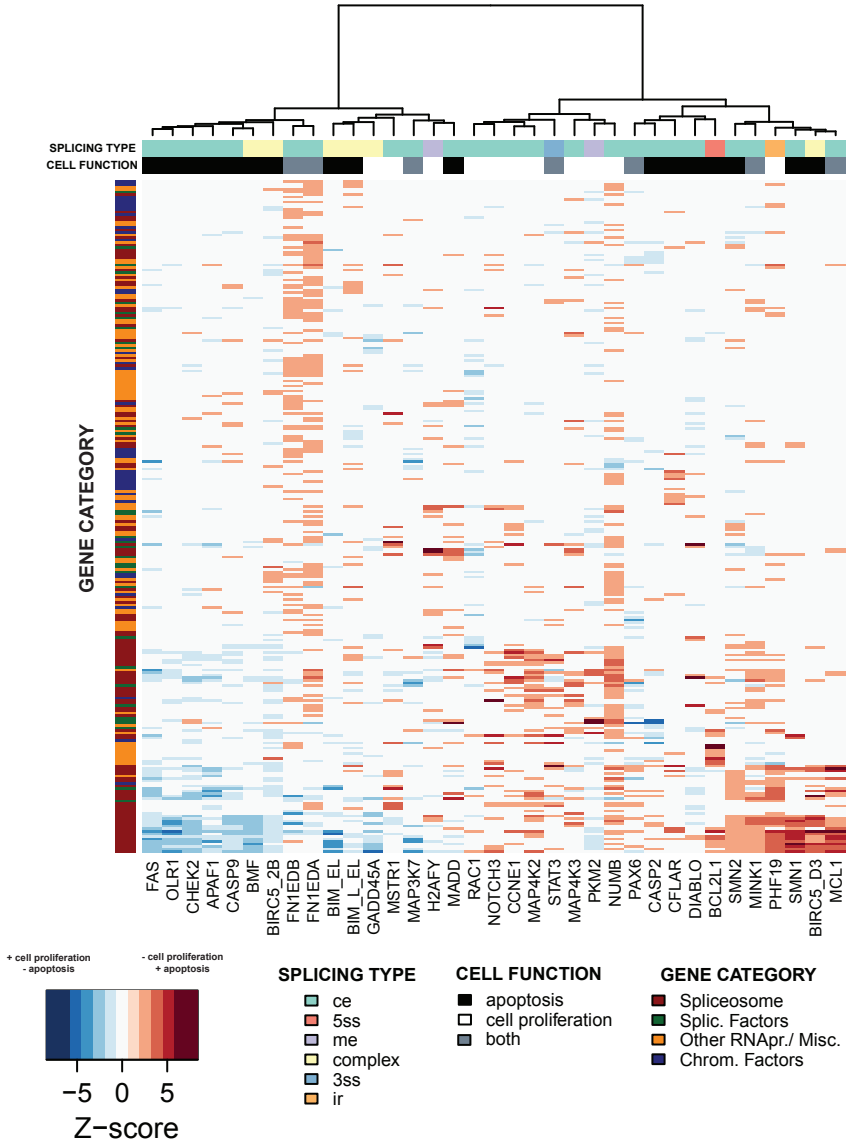

SPLICING TYPE
CELL FUNCTION
GENE CATEGORY

+ cell proliferation          - cell proliferation
- apoptosis                   + apoptosis

| | | |
|---|---|---|
| **SPLICING TYPE** | **CELL FUNCTION** | **GENE CATEGORY** |
| ce | apoptosis | Spliceosome |
| 5ss | cell proliferation | Splic. Factors |
| me | both | Other RNApr./ Misc. |
| complex | | Chrom. Factors |
| 3ss | | |
| ir | | |

−5   0   5
**Z-score**

**B**

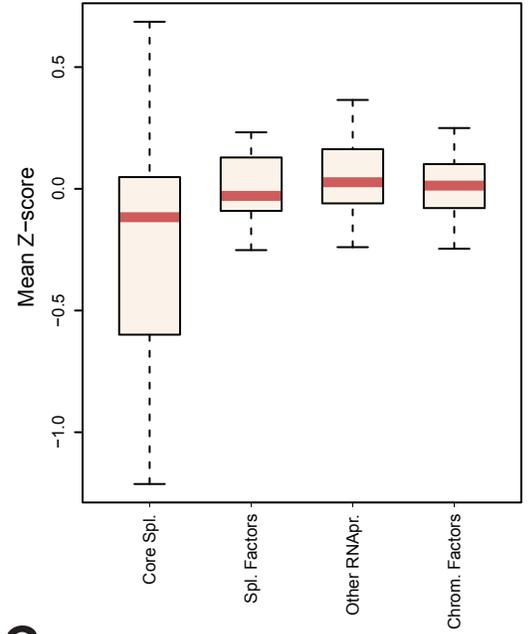

**C**

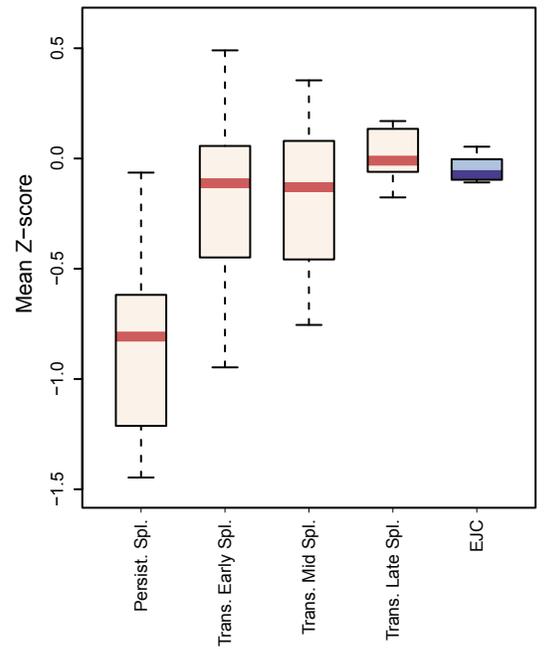

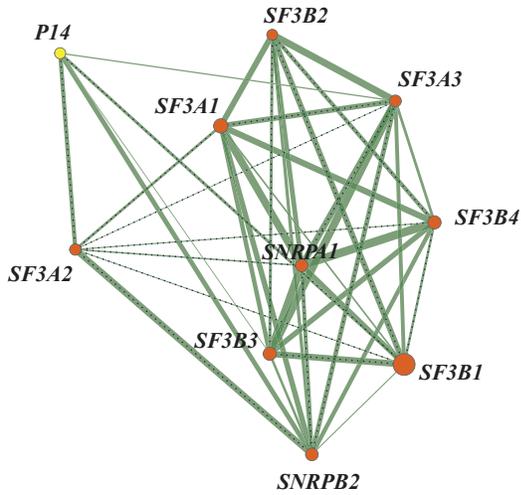

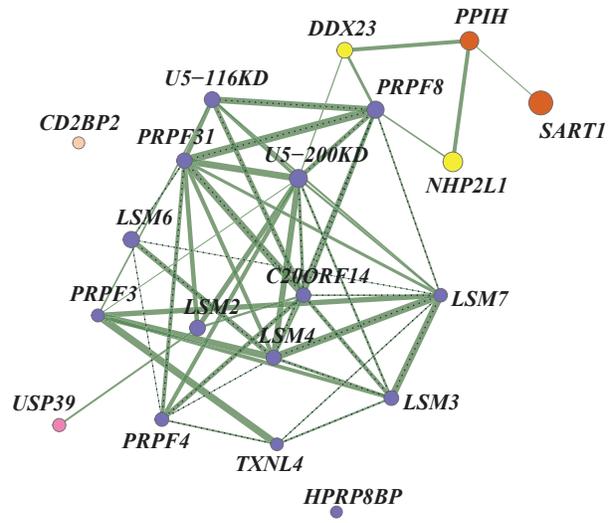

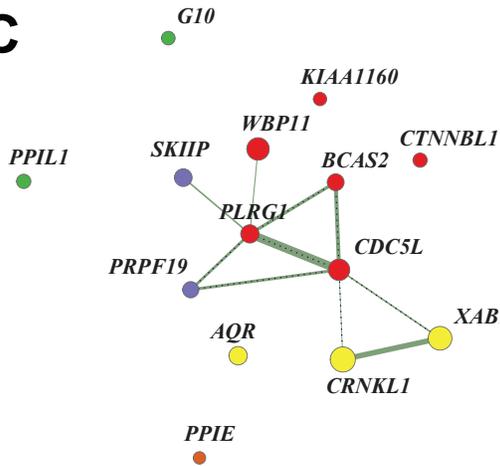

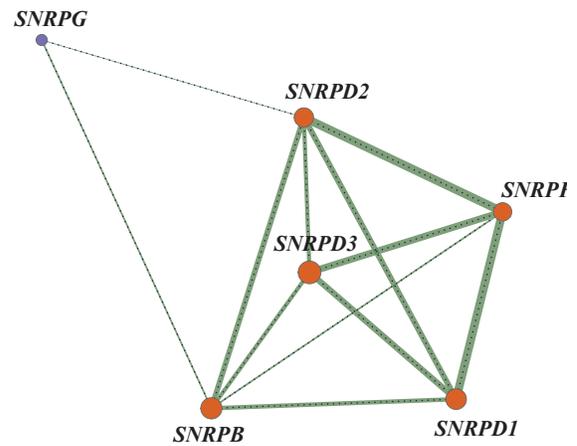

**E**

| | Persistent | Trans. Early | Trans. Mid. | Trans. Late |
|---|---|---|---|---|
| **Persistent** | SF3B4-SNRPA1<br>SF3B3-SNRPA1<br>SF3A1-SNRPA1<br>SF3A3-SF3B3<br>SF3A3-SNRPA1 | | | |
| **Trans. Early** | PPIH-SF3B2<br>PPIH-SNRPA1<br>PPIH-SF3B4<br>PRPF31-PRPF8<br>PPIH-SF3B1 | SNRPC-SNRP70<br>RBM17-SR140<br>IK-SMU1<br>PRPF31-C20ORF14<br>U2AF1-U2AF2 | | |
| **Trans. Mid.** | KIAA1604-P14<br>KIAA1604-SF3B1<br>PRPF19-U5200KD<br>XAB2-SNRPA1<br>XAB2-SF3B4 | KIAA1604-DDX23<br>CDC40-DDX23<br>AQR-DDX23<br>KIAA1604-PPIH<br>CDC40-PPIH | CDC5L-PLRG1<br>KIAA1604-CRNKL1<br>XAB2-CRNKL1<br>XAB2-CDC40<br>KIAA1604-XAB2 | |
| **Trans. Late** | C19ORF29-PRPF8<br>SLU7-P14<br>SLU7-SF3B2<br>C19ORF29-P14<br>SLU7-SF3B4 | LENG1-SMNDC1<br>PRPF18-LSM7<br>PRPF18-LSM4<br>SLU7-PPIH<br>LENG1-NHP2L1 | SLU7-CRNKL1<br>SLU7-KIAA1604<br>SLU7-XAB2<br>SLU7-CDC40<br>LENG1-AQR | SLU7-C19ORF29<br>DHX8-DDX41<br>KIAA0073-FLJ35382<br>DHX8-C19ORF29 |

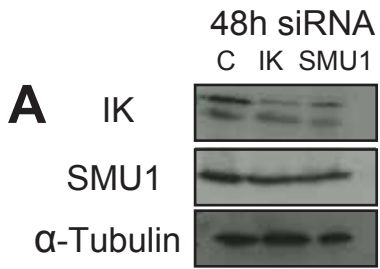

**48h siRNA**
C IK SMU1
IK
SMU1
α-Tubulin

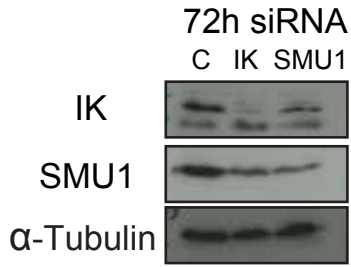

**72h siRNA**
C IK SMU1
IK
SMU1
α-Tubulin

**A**

**B**

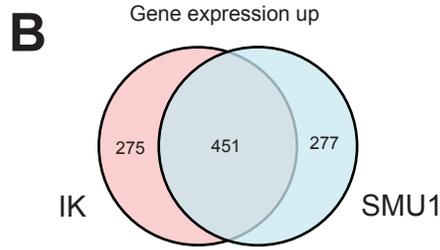

Gene expression up
IK 275 | 451 | 277 SMU1

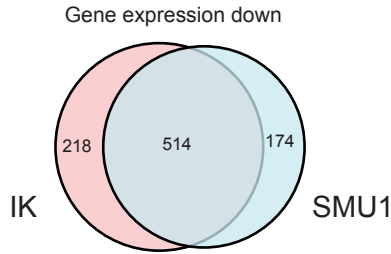

Gene expression down
IK 218 | 514 | 174 SMU1

**C**

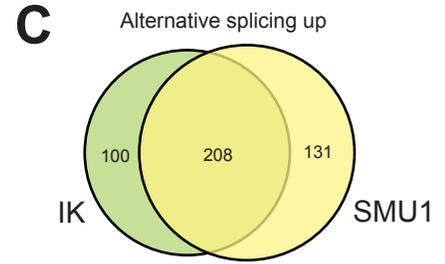

Alternative splicing up
IK 100 | 208 | 131 SMU1

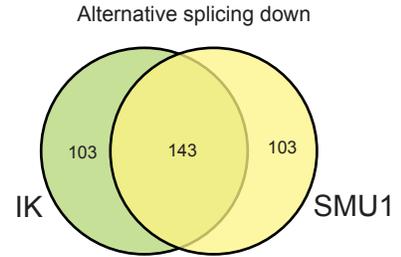

Alternative splicing down
IK 103 | 143 | 103 SMU1

**D**

### IK KD gene ontologies

**all genes (1450)**

| | p-value | # genes |
| --- | --- | --- |
| Cellular Growth and Proliferation | 8.49E-12 | 348 |
| Cell Death and Survival | 1.07E-10 | 342 |
| Gene Expression | 1.12E-09 | 212 |
| Cellular Movement | 2.90E-07 | 174 |
| Cell Cycle | 3.41E-06 | 135 |

**genes up (726)**

| | p-value | # genes |
| --- | --- | --- |
| Gene Expression | 2.49E-13 | 139 |
| Cell Death and Survival | 1.91E-05 | 175 |
| Cellular Growth and Proliferation | 1.93E-05 | 172 |
| Cellular Movement | 7.94E-05 | 78 |
| RNA Post-Transcriptional Modification | 8.18E-05 | 20 |

**genes down (732)**

| | p-value | # genes |
| --- | --- | --- |
| Cell-To-Cell Signaling and Interaction | 1.01E-07 | 91 |
| Cellular Growth and Proliferation | 1.15E-07 | 176 |
| Cellular Movement | 7.29E-07 | 106 |
| Cell Death and Survival | 1.72E-06 | 170 |
| Cell Morphology | 5.65E-06 | 102 |

**all exons (552)**

| | p-value | # genes |
| --- | --- | --- |
| RNA Post-Transcriptional Modification | 3.67E-11 | 49 |
| Gene Expression | 5.46E-09 | 195 |
| Cellular Assembly and Organization | 5.27E-05 | 171 |
| Cellular Function and Maintenance | 5.82E-05 | 148 |
| Cell Death and Survival | 5.95E-05 | 237 |

**exons up (308)**

| | p-value | # genes |
| --- | --- | --- |
| Cell Cycle | 2.04E-05 | 98 |
| Cell Morphology | 6.41E-05 | 114 |
| Gene Expression | 8.80E-05 | 132 |
| Cellular Assembly and Organization | 1.88E-04 | 125 |
| Cellular Function and Maintenance | 1.88E-04 | 111 |

**exons down (246)**

| | p-value | # genes |
| --- | --- | --- |
| RNA Post-Transcriptional Modification | 1.28E-09 | 26 |
| Gene Expression | 2.49E-08 | 72 |
| Cell Morphology | 4.15E-05 | 41 |
| Cell Cycle | 5.91E-05 | 59 |
| Cellular Assembly and Organization | 5.91E-05 | 54 |

### SMU1 KD gene ontologies

**all genes (1416)**

| | p-value | # genes |
| --- | --- | --- |
| Cell Death and Survival | 1.04E-08 | 311 |
| Cellular Growth and Proliferation | 1.52E-08 | 314 |
| Cell-To-Cell Signaling and Interaction | 1.37E-06 | 100 |
| Gene Expression | 1.59E-06 | 196 |
| Cellular Movement | 6.62E-06 | 164 |

**genes up (728)**

| | p-value | # genes |
| --- | --- | --- |
| Gene Expression | 1.23E-08 | 119 |
| Cellular Development | 1.63E-05 | 77 |
| Cell Death and Survival | 1.42E-04 | 154 |
| Gene Expression | 3.61E-04 | 68 |
| DNA Replication, Recombination, and Repair | 5.06E-04 | 26 |

**genes down (688)**

| | p-value | # genes |
| --- | --- | --- |
| Cellular Movement | 1.62E-09 | 102 |
| Cell-To-Cell Signaling and Interaction | 1.70E-08 | 66 |
| Cellular Growth and Proliferation | 1.44E-07 | 169 |
| Cell Death and Survival | 2.77E-06 | 159 |
| Cell Cycle | 4.94E-05 | 61 |

**all exons (583)**

| | p-value | # genes |
| --- | --- | --- |
| Gene Expression | 3.58E-09 | 202 |
| RNA Post-Transcriptional Modification | 4.60E-07 | 38 |
| Cellular Assembly and Organization | 2.25E-06 | 151 |
| Cellular Function and Maintenance | 2.25E-06 | 127 |
| Cell Cycle | 4.68E-06 | 133 |

**exons up (339)**

| | p-value | # genes |
| --- | --- | --- |
| Gene Expression | 2.76E-06 | 136 |
| Cellular Assembly and Organization | 7.41E-06 | 118 |
| Cellular Function and Maintenance | 1.87E-05 | 101 |
| Protein Synthesis | 2.75E-05 | 48 |
| Cell Cycle | 1.60E-04 | 92 |

**exons down (246)**

| | p-value | # genes |
| --- | --- | --- |
| RNA Post-Transcriptional Modification | 5.16E-07 | 21 |
| Gene Expression | 3.85E-05 | 65 |
| Cell Death and Survival | 5.15E-05 | 86 |
| Cellular Assembly and Organization | 1.33E-04 | 40 |
| Cellular Function and Maintenance | 1.33E-04 | 38 |

**E**

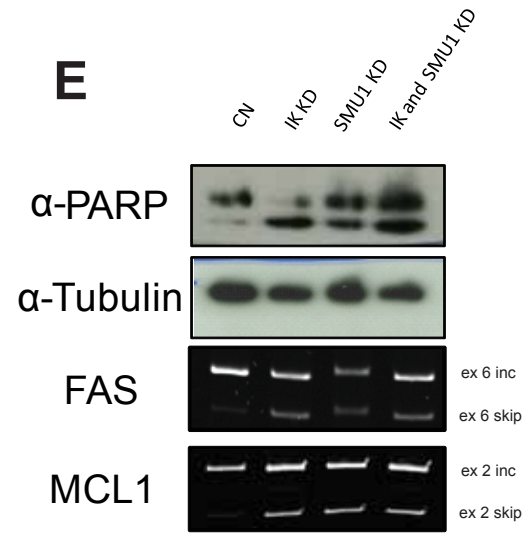

CN IK KD SMU1 KD IK and SMU1 KD

α-PARP
α-Tubulin
FAS — ex 6 inc / ex 6 skip
MCL1 — ex 2 inc / ex 2 skip

**F**

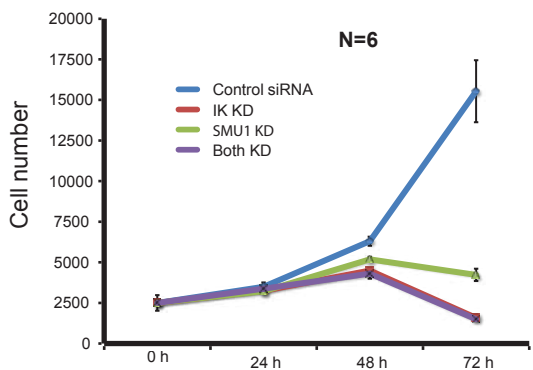

N=6

Cell number

— Control siRNA
— IK KD
— SMU1 KD
— Both KD

0 h 24 h 48 h 72 h

**G**

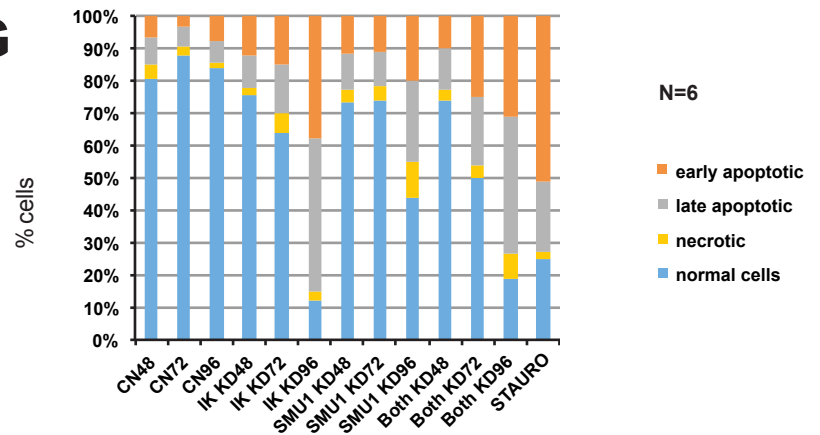

N=6

% cells

CN48 CN72 CN96 IK KD48 IK KD72 IK KD96 SMU1 KD48 SMU1 KD72 SMU1 KD96 Both KD48 Both KD72 Both KD96 STAURO

■ early apoptotic
■ late apoptotic
■ necrotic
■ normal cells

**A**

| n° | edge | n° | edge | n° | edge | n° | edge | n° | edge | n° | edge | n° | edge | n° | edge | n° | edge | n° | edge |
|---|---|---|---|---|---|---|---|---|---|---|---|---|---|---|---|---|---|---|---|
| 1 | SF3B1 SFRS10 | 11 | IK SF3B4 | 21 | SMNDC1 XAB2 | 31 | DDX10 MORG1 | 41 | DICER1 SFPQ | 51 | DDX31 NHP2L1 | 61 | BRD4 RNPC2 | 71 | SFRS1 TAF15 | 81 | CPSF5 FRG1 | 91 | DDX52 JMJD2B |
| 2 | DBR1 SKIP | 12 | NHP2L1 SNRPA1 | 22 | DHX30 BCAS2 | 32 | LSM6 MGC20398 | 42 | KIAA0073 SFPQ | 52 | DDX52 EXDSC4 | 62 | NONO RBM15 | 72 | DIS3 YBX1 | 82 | DEK PPIE | 92 | MGC13125 SMARCA4 |
| 3 | HNRPC U2AF1 | 13 | SMU1 SNRPA1 | 23 | BCAS2 MGC13125 | 33 | DDX14 SFRS11 | 43 | DH030 NONO | 53 | EXDSC4 USP39 | 63 | ILF3 SKIV2L2 | 73 | DH9I HNRPH2 | 83 | BUB3 SDCCAG10 | 93 | ACIN1 PRMT5 |
| 4 | NHP2L1 SF3B1 | 14 | NHP2L1 SNRPB2 | 24 | THOC2 WTAP | 34 | DDX62 SFRS2 | 44 | DH030 NONO | 54 | MGC22918 SKIV2L2 | 64 | MGC22918 SKIV2L2 | 74 | FNBP3 HNRPH2 | 84 | CRK7 PPIG | 94 | ACIN1 CTCF |
| 5 | RBM25 SF3B1 | 15 | D6S2864E USP39 | 25 | C02BP2 GTL3 | 35 | ELAVL1 SFRS10 | 45 | HNRPA0 NONO | 55 | FHX15 SNRPG | 65 | AOF2 DDX01 | 75 | FHX15 SNRPG | 85 | EZH2 PPIG | 95 | KHDRBS1 MORF4L1 |
| 6 | IK SF3B2 | 16 | C22ORF19 RY1 | 26 | P29 SRMP2 | 36 | P29 SRMP2 | 46 | LENG1 RBM35A | 56 | FHX15 SNRPG | 66 | DDX62 MFAP1 | 76 | DDX62 MFAP1 | 86 | CRK7 PPIG | | |
| 7 | NHP2L1 SF3A1 | 17 | NHP2L1 SNRPD1 | 27 | EXDSC4 SRMP2 | 37 | EXDSC4 C20ORF14 | 47 | BRD4 C20ORF14 | 57 | DDX62 LSM2 | 67 | HNRPA1 ZNF207 | 77 | FRG1 MFAP1 | 87 | ILF2 SETD1A | | |
| 8 | IK SF3A3 | 18 | NHP2L1 SNRPF | 28 | SF3B1 SNRPF | 38 | HNRPL SNRP70 | 48 | BRD4 PRPF4 | 58 | ASH2L PPM11G | 68 | DHX15 HSPC148 | 78 | FRG1 MFAP1 | 88 | HDAC4 SETD2 | | |
| 9 | NHP2L1 SF3A3 | 19 | DDX11 DDX01 | 29 | FRG1 SNRP70 | 39 | FRG1 SNRP70 | 49 | NOSIP PRPF4 | 59 | CTCF HNRPD | 69 | CTCF HNRPD | 79 | ILF2 NOSIP | 89 | CCDC55 KIN | | |
| 10 | NHP2L1 SF3B4 | 20 | SFRS1 XAB2 | 30 | SF3A1 SMU1 | 40 | FRG1 PRPF31 | 50 | PRPF31 PRPF4 | 60 | RBM25 SHARP | 70 | EED ILF3 | 80 | C20ORF14 FRG1 | 90 | DHX15 JMJD2B | | |

**B**     Cor = 0.773

**C**     Cor = -0.004

**D**     Cor = 0.899

**E**     Cor = 0.0659

**Functional splicing network reveals extensive regulatory potential of the core**

**Spliceosomal machinery**

Panagiotis Papasaikas[1,2, *], J. Ramón Tejedor[1,2,*], Luisa Vigevani[1,2]

and Juan Valcárcel[1-3]

# Supplemental Information.

## This file contains the following items:

- **Supplementary Figure legends**

- **Supplementary Table legends**

- **Supplementary Methods**

- **Supplementary References**

## Supplementary Figure legends.

**Figure S1. Consistent profiles of AS changes and limited cell death induced by the knockdown of core components using different siRNAs.** A. AS perturbation profiles obtained for depletion of SNRPG, P14 or SF3B1 using two different siRNAs . Robust correlation between conditions and siRNA sequences used for factor depletion are indicated. Positive Z-scores indicate changes towards exon inclusion, negative values indicate exon skipping. B. Knockdown efficiency for a subset of splicing factor components. Western blot analyses of protein depletion upon siRNA treatment for 6

spliceosome factors, for two of them two independent siRNA libraries, as indicated in A. Depletion efficiency was calculated by quantifying the levels of western blot signals using ImageQuant TL software (GE Healthcare Life sicences). C. Quantification of cell death upon depletion of core spliceosomal components or treatment with staurosporin. Data represents the percentage of normal, early apoptotic, late apoptotic or necrotic cells analyzed by flow cytometry sorting after staining with annexin V and propidium iodide. Values represent the mean and standard deviation of 3 independent biological replicates. D. Induction of apoptosis by staurosporine leads to AS changes distinct from those induced by the knock down or core splicing factors. Perturbation profiles were measured after 3 or 8 hours of staurosporin treatment (1 mM) and compared with the perturbation profiles obtained upon knockdown of IK, SMU1 or SF3B1 for 72 hours.

**Figure S2. Differences in knockdown conditions do not significantly alter correlation values within the network.** A. Perturbation profiles upon IK, SMU1 or SF3B1 knockdown for 24, 48 or 72 hours across the 35 splicing events used for network generation. B. Perturbation profiles upon IK or SMU1 depletion at 48 and 72 hours, respectively. C. Overall effects in AS changes across the 35 events after 24, 48 or 72 hours of siRNA treatment. Box plots represent the median and spread of Z-scores of the effects across different events upon the knockdown of IK, SMU1 or SF3B1 for the different timepoints. D. Correlation matrix and robust correlation values between IK or SMU1 knockdowns at different time points.

**Figure S3. Regulation of AS events involved in cell proliferation or apoptosis based on functional annotations.** A. Heatmap representation of the results obtained in the screen experiment based on functional annotations of the AS events. Blue or red colors indicate splicing changes towards more cell proliferation/less apoptosis or less cell proliferation/more apoptosis for a given knockdown condition, respectively. Data are clustered on both dimensions (similarity measure based on Pearson correlation, ward linkage). B and C. Extent and direction of AS changes upon knockdown of subsets of splicing factors. Box-plots represent the skewness and spread of the mean z-scores for AS splicing changes observed across the AS events analyzed for core and non-core splicing factors, other RNA processing factors and chromatin factors (A) or for factors that are either persistent or transient components of the spliceosome at different stages of assembly (B). EJC corresponds to components of the Exon Junction Complex.

**Figure S4. Sub-network topologies.** A. Detailed functional interactions of U2snRNP proteins. B. Detailed functional interactions among U4/5/6 tri-snRNP proteins. C. Functional interactions between PRP19 and Nineteen Complex components. D. Functional interactions between Sm core proteins. E. Top 5 functional network connections between persistent and different transient components of the spliceosome at different stages of assembly based on robust correlation values.

**Figure S5. Genome-wide regulation of AS by IK and SMU1 and links with cell proliferation and apoptosis**. A. Depletion of IK and SMU1 protein levels by siRNA-mediated knockdown at 48 or 72 hours. The proteins were detected by western blot

analysis under the indicated conditions. B. Overlap between gene expression changes upon knockdown of IK or SMU1 by RNAi. Upper panel: up-regulated genes. Lower panel: down-regulated genes. C. Overlap in AS changes upon knockdown of IK or SMU1 by RNAi. Upper panel: changes leading to up-regulation of alternative exons. Lower panel: changes leading to down-regulation of alternative exons. D. Gene ontology analysis for genes or exons up- or down-regulated upon IK or SMU1 knockdown using Ingenuity pathway tools. E. Induction of $\alpha$-PARP-cleavage, a feature of apoptotic cells, upon siRNA-mediated knockdown of IK, SMU1 or both. Western blots (two upper panels) were carried out using protein extracts of cells transfected with specific siRNAs or controls. $\beta$-tubulin was used as loading control. RT-PCR assays (lower two panels) were used to detect changes in AS of Fas/CD95 and MCL1 endogenous transcripts. F. IK and/or SMU1 knock reduce cell proliferation. HeLa cells were transfected in sixtuplicate with siRNAs against IK, SMU1 or the combination of both and cell numbers were determined using resazurin (which measures aerobic respiration) at different time points. Error bars represent standard deviations for a given siRNA condition. G. IK and/or SMU1 knock promote apoptosis. HeLa cells were transfected in sextuplicate with siRNAs against IK, SUM1 or both and IK and SMU1 were depleted by siRNA transfection, isolated at different time points and different cell populations analyzed by flow cytometry sorting after staining with annexin V and propidium iodide. The fraction of cells under different knockdown conditions is shown.

**Figure S6. Analysis of variable functional interactions in the AS network.** A. Principal components analysis of functional interactions extracted by subsets of AS events (see also Methods). Colors indicate the positions of the interactions in the Principal Component Analysis, which captures the different relative contributions of the 35 AS events in defining these connections. The top ten discriminatory AS events based on their loadings on the first two principal components are shown as grey arrows. The numbers identify the functional connections between splicing factors and are detailed in the table at the bottom. B and C. Representation of z-score values of AS changes upon knockdown of U2AF1 or HNRPC for AS events harboring (B) or not (C) composite HNRPC binding/3' splice site-like elements within 400 nucleotides of the alternative cassette exon boundaries. Robust correlation values are indicated within each panel. D and E. Representation of z-score values of AS changes upon knockdown of U2AF1 or HNRNPC for AS events harboring (D) or not (E) Alu elements within 0.5 kbp of the alternative cassette exon boundaries.

**Table S1. Information on AS events analyzed in this study.** The information includes gene symbol, id, exon sequence and genomic coordinates for the AS exons analyzed.

**Table S2. Information about the custom siRNA library utilized to knockdown splicing and chromatin factors.** Information includes siRNA pools catalog numbers, gene symbol, id and accession numbers for the targeted genes. It also includes information about the categories in which their protein products were classified according to two publications (Wahl et al., 2009 and Zhou et al., 2002), the classifications used in our network analysis and also information about their functions and known interactions.

**Table S3. Raw data of AS changes upon knockdown of splicing and chromatin factors.** Each sheet contains the data corresponding to one AS event. Data include information about the genes knocked down, the corresponding siRNA pools, quantification of alternative isoforms by RT-PCR and high-throughput capillary electrophoresis and percentage of exon inclusion (PSI index) for each of the three biological replicas as well as the median, standard deviation, p value and z-scores for the three values.

**Table S4 . Compendium of raw data of AS changes upon knockdown of splicing and chromatin factors (classified according to factor classes), upon knockdown of P14, SF3B1 and SNRPG with two independent siRNA libraries, upon knockdown of IK-SMU1 core splicing factors at different time points or in different cell lines, upon treatment with drugs targeting the spliceosome and upon pharmacological treatments involved in iron homeostasis modulation (data from Tejedor et al., accompanying manuscript).**

**Table S5. Robust correlation values and glasso inverse covariance estimates for the complete network edge-list.**

**Table S6. Oligonucleotide sequences and information about amplicons utilized in semi quantitative RT-PCR (sheet 1) assays or real time PCR analysis (sheet 2) used in this study**

**Detailed information and references for the Splicing events analyzed can be found at the following link:**

https://s3.amazonaws.com/SplicingNet/ASEs.zip

The files include a schematic representation for every AS event; information regarding the functional role of the gene, as retrieved from uniprot / genecards database; the protein domain affected by the AS event; the functional effect of each of the isoforms towards cell proliferation or apoptosis; and 102 references consulted for the selection.

**Supplementary Methods**

**Cell lines**

HeLa CCL-2 cells and HEK293 cells were purchased from the American Type Culture Collection (ATCC). Cells were cultured in Glutamax Dulbecco's modified Eagle's medium (Gibco, Life Technologies) supplemented with 10% fetal bovine serum (Gibco, Life Technologies) and penicillin/streptomycin antibiotics (penicillin 500 u/ml; streptomycin 0.5 mg/ml, Life Technologies). Cell culture was performed in cell culture dishes in a humidified incubator at 37$^o$C under 5% CO2.

**Drug Treatments**

Pharmacological treatments with splicing arresting drugs were performed as described in Corrionero et al. (2011). HeLa cells were treated for 3 hours with Spliceostatin A 260 nM or 8 hours with Meayamycin 20 nM or TG003 10 μM. RNA was isolated using the Maxwell 16 LEV simplyRNA kit (Promega) and RT PCRs for the 35 splicing events were performed as described above.

**Cell proliferation assays**

HeLa cells were forward-transfected with siRNAs targeting IK, SMU1 or both genes by plating 2500 cells over a mixture containing siRNA-RNAiMAX lipofectamine complexes, as previously described. Indirect estimations of cell proliferation were obtained 24, 48 and 72 hours post transfection by treating the cells for 4h with resazurin (5 µM) prior to the fluorescence measurements (544 nm excitation and 590 nm emission wavelengths). Plate values were obtained with an Infinite 200 PRO series multiplate reader (TECAN).

**Cell apoptosis assays**

Hela cells were transfected with siRNAs against IK, SMU1 or both genes, as described above. Cells were trypsinized and collected for analysis 72 hours post transfection. Pellets were washed once with complete DMEM medium (10% serum plus antibiotics), and three times with PBS 1X. Cells were then resuspended in 200 µl Binding buffer provided by the Annexin V-FITC Apoptosis Detection Kit (eBiosciences), resulting in a final density of $4 \times 10^5$ cells/ml. 5 µl of Annexin V-FITC was added to each sample followed by 10 min incubation at room temperature in dark conditions. Cells were washed once with 500 µl binding buffer and incubated with 10 µl propidium iodide (20 µg/ml) for 5 minutes. FACS analysis was performed using a FACSCalibur flow cytometry system (BD Biosciences).

**Western blots**

Protein extraction was performed in RIPA buffer and protein abundance was estimated by Bradford assay. Thirty to Forty µg of total protein were fractionated by electrophoresis in 10% Bis-Tris polyacrilamide gels and transferred to PVDF membranes. Primary antibodies used in this study are listed below: IK (rabbit polyclonal, Origene TA308401), SMU1 (mouse monoclonal, Santa Cruz Biotech, sc-100896 JS-12), PARP (rabbit polyclonal, cell signaling, 9542S), β-Tubulin (mouse monoclonal, Sigma-Aldrich, T4026), SF3B1 (rabbit polyclonal, Abcam, ab39578), SNRPG (rabbit polyclonal, Abcam, ab111194), SLU7 (rabbit polyclonal, gift from Robin Reed), U2AF65 (mouse monoclonal, MC3 clone), SRSF1 (mouse monoclonal, gift from Adrian Krainer), BRR2 (rabbit polyclonal, Sigma, -HPA029321)

**Real time qPCR**

First strand cDNA synthesis was set up with 500 ng of RNA, 50 pmol of oligo-dT (Sigma-Aldrich), 75 ng of random primers (Life Technologies), and superscript III reverse transcriptase (Life Technologies) in 20 µl final volume, following the manufacturer's instructions. Quantitative PCR amplification was carried out using 1 µl of 1:2 to 1:4 diluted cDNA with 5 µl of 2X SYBR Green Master Mix (Roche) and 4 pmol of specific primer pairs in a final volume of 10 µl in 384 well-white microtiter plates, Roche). qPCR mixes were analyzed in triplicates in a Light Cycler 480 system (Roche) and fold change ratios were calculated according to the Pfaffl method (Pfaffl, 2001). Primers sequences used in this study are indicated in Table S6.

**Affymetrix arrays**

HeLa cells were transfected with siRNAs against IK, SMU1 or RBM6 in biological triplicate as described above or –for RBM6- as described in Bechara et al., 2013. RNA was isolated using the RNeasy minikit (Quiagen). RNA quality was estimated by Agilent Bioanalyzer nano assay and samples were hybridized to GeneChip Human Exon 1.0 ST Array (Affymetrix) by Genosplice Technology. Array analysis was performed by Genosplice Technology and the data (gene expression, alternative splicing changes and gene ontologies upon IK, SMU1 knockdown conditions) is available at: https://s3.amazonaws.com/SplicingNet/HJAY_arrays.zip and at GEO database with the accession number GSE56605 .

**Glasso-based Network Reconstruction from Robust Correlation Estimates**

Here we describe the process of deriving an undirected network that models functional interconnections of splicing perturbations based on the effects of gene KDs/cell-perturbing stimuli on ASEs. The main steps are:

1. Data preprocessing and reduction. This involves:

    1.1. Removal of sparse variables and imputation of missing values.

    1.2. Removal of uninformative events.

    1.3. Data scaling (standardization).

2. Robust sample covariance estimation.

3. Network reconstruction based on the graphical lasso (glasso) model selection algorithm.

In the next sections we describe the basis and algorithmic details for each of the above steps.

*1.      Data preprocessing and reduction.*

*1.1 Removal of sparse variables and imputation of missing values*

Estimation of the sample covariance matrix requires complete data. However the $p$ **x** $n$ matrix that contains the measurements of the $p$ variables (genes) in $n$ conditions (ASEs) includes missing (*NA*) values. In order to overcome this problem, first sparse variables (*#NA values > n/2*) are removed. The remaining missing values are subsequently filled-in by applying *k*-nearest-neighbor based imputation (Hastie et al., 1999, R function *impute* available as part of the bioconductor project at http://www.bioconductor.org/). For the purposes of this work we set *k=ceiling(0.25 sqrt(p))*.

*1.2 Removal of uninformative ASEs.*

ASEs not affected to a significant degree (*median absolute (ΔPSI)<1*) by the KDs offer little information on the genes' covariance and can potentially introduce noise and are thus discarded from subsequent analyses.

*1.3 Data Scaling.*

Prior to sample covariance estimation the data matrix is standardized along the gene dimensions. Standardization of the events ensures that only the shape and not the magnitude of the fluctuations is taken into account for covariance estimation and it is equivalent to using correlation in lieu of covariance in all downstream analyses.

*2. Robust sample correlation estimation.*

When outliers are present they dominate the correlation estimates. In order to overcome the sensitivity of sample correlation to outliers we derive robust correlation estimates based on a weighting algorithm that attempts to discriminate between technical outliers and reliable influential measurements.

In particular, starting with an $M = p$ x $n$ dataset of $p$ standardized variables and $n$ samples we wish to derive a robust correlation $p$ x $p$ matrix. We calculate the influence of a sample $u$ in the estimation of the correlation of two variables $X_a$, $X_b$ using the deleted residual distance:

$$D_u(X_a, X_b) = |p(X_a, X_b) - p_{-u}(X_a, X_b)|$$

Where $p(X_a, X_b)$ and $p_{-u}(X_a, X_b)$ are the Pearson's correlation estimates before or after removing the observations $M_{au}$, $M_{bu}$ coming from sample $u$.

A measure of the reliability $R_{a,b}^u$ of these measurements for the estimation of $p(X_a, X_b)$ given all the observed data is then based on their cumulative influence on high correlates of $X_a$ and $X_b$:

$$R_{a,b}^u = 1 / \frac{\sum_{i=1}^{p} D_u(X_a, X_i)\mathbb{I}(a,i) + \sum_{i=1}^{p} D_u(X_b, X_i)\mathbb{I}(b,i)}{\sum_{i=1}^{p} \mathbb{I}(a,i) + \sum_{i=1}^{p} \mathbb{I}(b,i)}$$

where $\mathbb{I}(a,i)$ is the indicator function:

$$\mathbb{I}(a,i) := \begin{cases} 1 \text{ if } i \neq a \text{ } AND \text{ } |p(X_a, X_i)| > T \\ \qquad 0 \text{ otherwise} \end{cases}$$

$T$ being a minimum correlation threshold for considering a pair of variables.

The final robust correlation estimates are finally calculated as the weighted Pearson's correlation $p(X_a, X_b;w)$ where $w$ is a vector of weights of length $n$ containing the values $R_{a,b}^u$ for every observation $u$.

Intuitively, if an observation $M_{au}$ distorts correlation estimates among all highly correlated partners of $X_a$ it is considered an outlier and is down-weighted. On the other hand if including this observation yields similar correlation estimates for most of these pairs it is considered reliable even if its influence on the individual estimate of $p(X_a, X_b)$ is high. The advantage of this procedure is that it derives weights for individual measurements (as opposed to a single weight for every sample) while taking into account the complete dataset.

The algorithm can be used iteratively, however in our experience the estimates converge within 1 to 2 iterations. R code for implementing this algorithm is available upon request. Examples of robust correlation estimates and regression based on this metric and taken from our dataset are shown below:

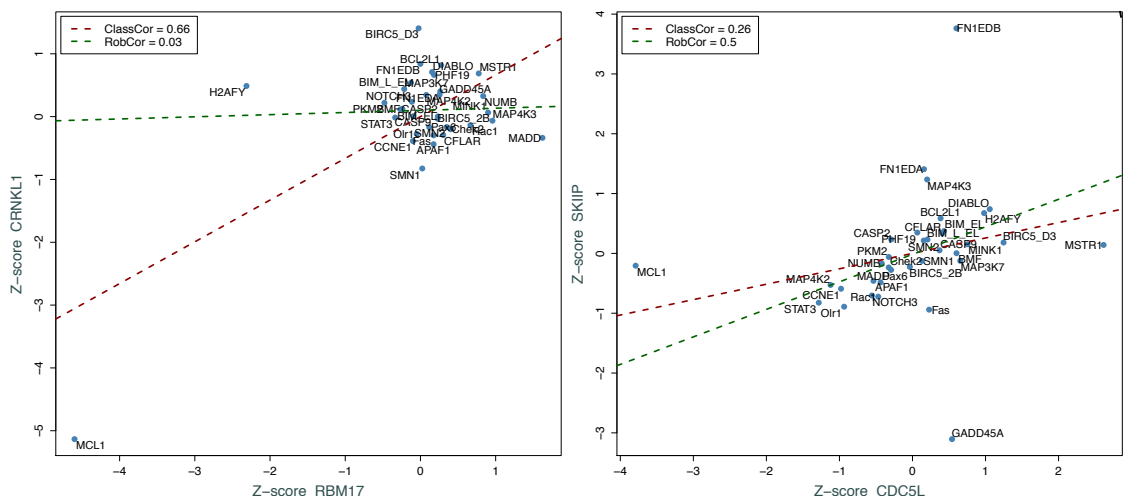

3. *Graphical model selection for Network Reconstruction using Graphical Lasso*

Undirected graphical models (UGMs) such as Markov random fields (MRFs) represent real-world networks and attempt to capture important structural and functional aspects of the network in the graph structure. The graph structure encodes conditional independence assumptions between variables corresponding to nodes of the network. The problem of recovering the structure of the graph is known as model selection or covariance estimation.

In particular, let $G=(V,E)$ be an undirected graph on $p=|V|$ nodes. Given $n$ independent, identically distributed (i.i.d) samples of $X=(X1,…,Xp)$, we wish to identify the underlying graph structure. We restrict the analysis to Gaussian MRFs where the model assumes that the observations are generated from a multivariate Gaussian distribution $N(\mu,\Sigma)$. Based on the observation that in the Gaussian setting, zero components of the inverse covariance matrix $\Sigma^{-1}$ correspond to conditional independencies given the other variables, different approaches have been proposed in order to estimate $\Sigma^{-1}$.

Graphical lasso (gLasso) provides an attractive solution to the problem of covariance estimation for undirected models, when graph sparsity is a goal (Friedman et al., 2008). The gLasso algorithm has the advantage of being consistent, i.e. in the presence of infinite samples its parameter estimates will be arbitrarily close to the true estimates with probability 1.

$L_1$ regularization (lasso) is a smooth form of subset selection for achieving sparsity. In the case of gLasso the model is constructed by optimizing the log-likelihood function:

$$\log \det \Theta - tr(S\Theta) - r\|\Theta\|_1$$

Where $\Theta$ is an estimate for the inverse covariance matrix $\Sigma^{-1}$, $S$ is the empirical covariance matrix of the data, $||\Theta||_1$ is the $L1$ norm i.e the sum of the absolute values of all elements in $\Theta^{-1}$, and $r$ is a regularization parameter (in this case selected based on estimates of the FDR, see below). The solution to the glasso optimization is convex and can be obtained using coordinate descent.

The glasso process for network reconstruction was implemented using the glasso R package by Jerome Friedman, Trevor Hastie and Rob Tibshirani (http://statweb.stanford.edu/~tibs/glasso/).

4. **Partition of Network into modules.**

Network modules or communities can be defined loosely as sets of nodes with a more dense connection pattern among their members than between their members and the remainder of the. Module detection in real-world graphs is a problem of considerable practical interest as they often capture meaningful functional groupings.

Typically community structure detection methods try to maximize modularity, a measure of the overall quality of a certain network partition in terms of the identified modules (Newman, 2006). In particular, for a network partition $p$, the network modularity $M(p)$ is defined as:

$$M(p) = \sum_{k=1}^{m} \left[\frac{l_k}{L} - \left(\frac{d_k}{2L}\right)^2\right]$$

where $m$ is the number of modules in $p$, $l_k$ is the number of connections within module $k$, $L$ is the total number of network connections and $d_k$ is the sum of the degrees of the nodes in module $k$.

Here, we identify modules of genes that exhibit similar perturbation profiles among the assayed events by maximizing the network's modularity using the greedy community detection algorithm (Clauset et al., 2004) implemented in the *fastgreedy.community* function of the igraph package (Csardi and Nepusz, 2006, http://igraph.sourceforge.net/doc/R/fastgreedy.community.html).